\documentclass[openbib,12pt]{article}

\newcommand{\utwi}[1]{\mbox{\boldmath $ #1$}}

\usepackage{booktabs,dcolumn}
\usepackage{psfrag,graphicx}
\usepackage{eso-pic,graphicx}
\usepackage{amsmath}
\usepackage{graphicx}
\usepackage{pdflscape}
\usepackage{longtable}
\usepackage{epstopdf}

\newcolumntype{z}[1]{D{.}{.}{#1}}

\newcommand{\cred}{\textcolor{red}}
\newcommand{\cb}{\textcolor{blue}}
\date{}

\begin{document}

\title{
\begin{center} {\Large \bf Bayesian Semi-parametric Realized-CARE Models for Tail Risk Forecasting Incorporating Realized Measures} \end{center}}
\author{Richard Gerlach, Chao Wang\\
Discipline of Business Analytics, Business School, \\The University of Sydney, Australia.}

\date{} \maketitle

\begin{abstract}
\noindent
A new model framework called Realized Conditional Autoregressive Expectile (Realized-CARE) is proposed, through incorporating a measurement
equation into the conventional CARE model, in a manner analogous to the Realized-GARCH model. Competing realized measures
(e.g. Realized Variance and Realized Range) are employed as the dependent variable in the measurement equation and to drive expectile dynamics.
The measurement equation here models the contemporaneous dependence between the realized measure and the latent conditional
expectile. We also propose employing the quantile loss function as the target criterion, instead of the conventional violation rate,
during the expectile level grid search. For the proposed model, the usual search procedure and asymmetric least squares (ALS) optimization to
estimate the expectile level and CARE parameters proves challenging and often fails to convergence. We incorporate a fast random walk Metropolis stochastic search method, combined with a more targeted grid search procedure,
to allow reasonably fast and improved accuracy in estimation of this level and the associated model parameters.
Given the convergence issue, Bayesian adaptive Markov Chain Monte Carlo methods are proposed for estimation, whilst their properties are
assessed and compared with ALS via a simulation study. In a real forecasting study applied to 7 market indices and
2 individual asset returns, compared to the original CARE, the parametric GARCH and Realized-GARCH models, one-day-ahead Value-at-Risk and
Expected Shortfall forecasting results favor the proposed Realized-CARE model, especially when incorporating
the Realized Range and the sub-sampled Realized Range as the realized measure in the model.

\vspace{0.5cm}

\end{abstract}

\newpage
\pagenumbering{arabic}

{\centering
\section{\normalsize INTRODUCTION}\label{introduction_sec}
\par
}
\noindent
In recent decades, quantitative financial risk measurement has provided a fundamental toolkit for investment decisions, capital allocation
and external regulation. Value-at-Risk (VaR) and Expected Shortfall (ES) are tail risk measures that are employed, as part of this toolkit,
to help measure and control financial risk. VaR represents the market risk as one number, a quantile of the risk distribution, and has
become a standard measurement for capital allocation and risk management, since it was proposed in 1993. However, VaR has been
criticized because it cannot measure the expected loss for violating returns and is not mathematically coherent, in that it can
favour non-diversification. ES, proposed by Artzner \emph{et al.} (1997, 1999), gives the expected loss, conditional on returns exceeding
a VaR threshold, and is a coherent measure, thus in recent years it has become more widely employed for tail risk measurement and is now recommended in the Basel Capital Accord.

Volatility estimation can play a key role in calculating accurate VaR or ES forecasts. Since the introduction of the
Auto-Regressive Conditionally Heteroskedastic (ARCH) model of Engle (1982) and the generalized (G)ARCH of Bollerslev (1986),
both employing squared returns as model input, very many different volatility measures and models have been developed.
Since Parkinson (1980) and Garman and Klass (1980) proposed the daily high-low range as a more efficient volatility estimator
compared to the daily squared return, the availability of high frequency intra-day data has generated several popular and efficient
realized measures, including Realized Variance (RV): Andersen and Bollerslev (1998), Andersen \emph{et al.} (2003);
and Realized Range (RR): Martens and van Dijk (2007), Christensen and Podolskij (2007). In
order to further deal with the well-known, inherent micro-structure noise accompanying high frequency volatility measures, Zhang, Mykland
and A\"{i}t-Sahalia (2005) and Martens and van Dijk (2007) designed the sub-sampling and scaling processes, respectively, aiming to provide
smoother and more efficient realized measures. In this paper the method of sub-sampling is extended to apply to the realized range measure.

Hansen \emph{et al.} (2011) extended the parametric GARCH model framework by proposing the Realized-GARCH (Re-GARCH), adding a measurement
equation that contemporaneously links unobserved volatility with a realized measure. Gerlach and Wang (2015) extended the Re-GARCH model
through employing RR as the realized measure and illustrated that the proposed Re-GARCH-RR framework can generate
more accurate and efficient volatility, VaR and ES forecasts compared to traditional GARCH and Re-GARCH models. Hansen and Huang (2016) recently
extended the parametric Re-GARCH framework to include multiple realized measures. However,
the tail-risk forecast performance of these parametric volatility models heavily depends on the choice of error distribution. A
semi-parametric model that directly estimates quantiles and expectiles, and
implicitly ES, called the Conditional Autoregressive Expectile (CARE) model is proposed by Taylor (2008). The relevant expectile
can be estimated via Asymmetric Least Squares (ALS), which is transformed to be an estimate of ES through a connection
discovered by Newey and Powell (1987). Gerlach, Chen and Lin (2012) developed the non-linear family of CARE models and an
associated Bayesian estimation framework. Further, Gerlach and Chen (2016) extended CARE type models through employing daily
high-low range as input, whilst Gerlach, Walpole and Wang (2016) proposed a CARE-X framework; again finding that allowing RR to drive the
model dynamics led to more accurate tail risk forecasts.

In this paper, a Realized Conditional Autoregressive Expectile (Re-CARE) framework is proposed, which is roughly analogous to
the Re-GARCH framework and is close to a semi-parametric Realized GARCH model. The Re-CARE incorporates the CARE model, but
adds a measurement equation that links the latent conditional expectile with the realized measure. The work in Gerlach and Chen (2016)
allows a likelihood formulation for CARE models, giving an MLE that is equivalent to the ALS estimator. A standard parametric
assumption on the errors of the Re-CARE measurement equation allows this formulation to be extended, permitting an Re-CARE likelihood to be
developed and an ML estimator to be explored. Further, an adaptive Bayesian MCMC algorithm is developed as a competitor for the Re-CARE model.
To evaluate the performance of the proposed Re-CARE models, employing the range and various realized measures as inputs, the accuracy of VaR
and ES forecasts will be assessed and compared with competitors such as the CARE, GARCH and Re-GARCH models.

The CARE model includes a nuisance parameter, currently not estimable by standard methods, for which Taylor (2008) employed
a grid search estimator that optimized the sample violation rate. This paper extends that approach in three ways. First, the quantile loss
function is proposed as a more suitable optimization criterion. Second, a two-step search method, consisting of a coarse grid, followed
by a refined grid search, is proposed as an alternative, helping to reduce the computing time in estimating this parameter, whilst
maintaining an equivalent level of accuracy. Third, a fast, efficient random walk Metropolis method is proposed to perform the
optimization, which corrects a convergence issue for this parameter in the proposed model.

The paper is organized as follows: Section \ref{realized_measure_section} reviews some realized measures and proposes the sub-sampled RR.
Expectiles and their connection with existing CARE type models, as well as a review of Re-GARCH type models comprises
Section\ref{expectile_care_section}. Section \ref{model_section} proposes the Realized-CARE class of model; the associated likelihood and
the adaptive Bayesian MCMC algorithm for parameter estimation are presented in Section \ref{beyesian_estimation_section}.
The simulation and empirical studies are discussed in Section \ref{simulation_section} and Section \ref{data_empirical_section} respectively.
Section \ref{conclusion_section} concludes the paper and discusses future work.

{\centering
\section{\normalsize REALIZED MEASURES}\label{realized_measure_section}
\par
}
\noindent
This section provide a review of some realized measures and proposes the sub-sampled Realized Range.

For day $t$, representing the daily high, low and closing prices as $H_{t}$, $L_{t}$ and $C_{t}$, the most commonly used daily log return is:
\begin{equation}\label{return_def}
r_t= \text{log}(C_t)-\text{log}(C_{t-1}) \nonumber
\end{equation}
where $r_t^2$ is the associated volatility estimator. The high-low range (squared), proposed by Parkinson (1980), proved to be a much more
efficient volatility estimator than $r_t^2$, based on the range distribution theory (see e.g. Feller, 1951):
\begin{equation}\label{range_def}
Ra_{t}^{2}=\frac {(\text{log}H_{t}-\text{log}L_{t})^2} {4\log2} \nonumber
\end{equation}
where $4 \text{log}(2)$  scales Ra to be an approximately unbiased volatility estimator. Several other range-based estimators, e.g.
Garman and Klass (1980); Rogers and Satchell (1991); Yang and Zhang (2000) were subsequently proposed; see Moln{\'a}r (2012) for a
full review regarding their properties. The range allowing for overnight price jumps is proposed in Gerlach and Chen (2015):
\begin{equation}
RaO_{t}= \text{log} \big( \text{max}(H_{t},C_{t-1})\big)-\text{log} \big( \text{min}(L_{t}, C_{t-1})
\end{equation}
where again the associated volatility estimator squares $RaO_t$, then divides by $4 \text{log}(2)$.

If each day $t$ is divided into $N$ equally sized intervals of length $\Delta$, subscripted by $\Theta= {0, 1, 2, ... , N}$, several
high frequency volatility measures can be calculated. For day $t$, denote the $i$-$th$ interval closing price as $P_{t-1+i \triangle}$ and
$H_{t,i}=\text{sup}_{(i-1) \triangle<j< i \triangle}P_{t-1+j}$ and $L_{t,i}=\text{inf}_{(i-1) \triangle<j<i \triangle}{P_{t-1+j}}$ as the high and
low prices during this time interval. Then RV is proposed by Andersen and Bollerslev (1998) as:
\begin{equation}\label{rv_def2}
RV_{t}^{\triangle}=\sum_{i=1}^{N} [log(P_{t-1+i \triangle})-log(P_{t-1+(i-1)\triangle})]^{2}
\end{equation}
Martens and van Dijk (2007) and Christensen and Podolskij (2007) developed the Realized Range, which sums the squared intra-period
ranges:
\begin{equation}\label{rrv_def}
RR_{t}^{\triangle}= \frac {\sum_{i=1}^{N}(\text{log}H_{t,i}-\text{log}L_{t,i})^2}{4\log2}
\end{equation}

Through theoretical derivation and simulation, Martijns and van Dijk (2007) show that RR is a competitive, and sometimes more efficient,
volatility estimator than RV under some micro-structure conditions and levels. Gerlach and Wang (2015) confirm that RR can provide extra
efficiency in empirical tail risk forecasting, when employed as the measurement equation variable in an Re-GARCH model.
To further reduce the effect
of microstructure noise, Martens and van Dijk (2007) presented a scaling process, as in Equations (\ref{rv_scale}) and (\ref{rrv_scale}).
\begin{eqnarray}\label{rv_scale}
RV_{S,t}^{\triangle}= \frac {\sum_{l=1}^{q}RV_{t-l}}{\sum_{l=1}^{q}RV_{t-l}^{\triangle}}RV_{t}^{\triangle},
\end{eqnarray}
\begin{eqnarray}\label{rrv_scale}
RR_{S,t}^{\triangle}= \frac {\sum_{l=1}^{q}RR_{t-l}}{\sum_{l=1}^{q}RR_{t-l}^{\triangle}}RR_{t}^{\triangle},
\end{eqnarray}
\noindent
where $RV_{t}$ and $RR_{t}$ represent the daily squared return and squared range on day $t$,, respectively. This scaling process is
inspired by the fact that the daily squared return and range are each less affected by micro-structure noise than their high frequency
counterparts, thus can be used to scale and smooth RV and RR, creating less micro-structure sensitive measures.

Further, Zhang, Mykland and A\"{i}t-Sahalia (2005) proposed a sub-sampling process, also to deal with micro-structure effects.
For day $t$, $N$ equally sized samples are grouped into $M$ non-overlapping subsets $\Theta^{(m)}$ with size $N/M=n_{k}$, which means:
\begin{equation}
\Theta= \bigcup_{m=1}^{M} \Theta^{(m)}, \; \text{where} \; \Theta^{(k)}  \cap \Theta^{(l)} = \emptyset, \;
\text{when}  \;  k \neq l.  \nonumber
\end{equation}
Then sub-sampling will be implemented on the subsets $\Theta^{i}$ with $n_{k}$ interval:
\begin{equation}
\Theta^{i}= {i, i+n_k,...,i+n_k(M-2), i+n_k(M-1)}, \; \text{where} \;  i= {0,1,2...,n_k-1}.  \nonumber
\end{equation}

Representing the log closing price at the $i$-$th$ interval of day $t$ as $C_{t,i}=P_{t-1+i\triangle}$, the RV with the subsets
$\Theta^{i}$ is:
\begin{equation}
RV_{i}= \sum_{m=1}^{M} (C_{t,i+n_{k}m}-C_{t,i+n_{k}(m-1)})^{2}; \; \text{where} \; i= {0,1,2...,n_k-1}.  \nonumber
\end{equation}

We have the $T/M$ RV with $T/N$ sub-sampling as (supposing there are $T$ minutes per trading day):

\begin{equation}
RV_{T/M,T/N}= \frac{\sum_{i=0}^{n_k-1} RV_{i} } {n_k},
\end{equation}

Then, denoting the high and low prices during the interval $i+n_{k}(m-1)$ and $i+n_{k}m$ as
$H_{t,i}=\text{sup}_{(i+n_{k}(m-1))\triangle<j<(i+n_{k}m) \triangle}P_{t-1+j}$ and
$L_{t,i}=\text{inf}_{(i+n_{k}(m-1))\triangle<j<(i+n_{k}m) \triangle}{P_{t-1+j}}$ respectively, we propose the $T/M$ RR with $T/N$ sub-sampling
as:
\begin{equation}
RR_{i}= \sum_{m=1}^{M} (H_{t,i}-L_{t,i})^2; \; \text{where} \; i= {0,1,2...,n_k-1}.
\end{equation}
\begin{equation}
RR_{T/M,T/N}= \frac{\sum_{i=0}^{n_k-1} RR_{i} } {4 \text{log}2 n_k},
\end{equation}

For example, the 5 mins RV and RR with 1 min subsampling can be calculated as below respectively:
\begin{eqnarray}  \nonumber
&&RV_{5,1,0}=(\text{log} C_{t5}-\text{log} C_{t0})^2+(\text{log} C_{t10}-\text{log} C_{t5})^2+... \\  \nonumber
&&RV_{5,1,1}=(\text{log}C_{t6}-\text{log} C_{t1})^2+(\text{log} C_{t11}-\text{log} C_{t6})^2+... \\ \nonumber
&&RV_{5,1}=\frac{\sum_{i=0}^{4}RV_{5,1,i}} {5}  \nonumber
\end{eqnarray}
\begin{eqnarray}  \nonumber
&&RR_{5,1,0}=(\text{log}H_{t0\leq t \leq t5}-\text{log} L_{t0\leq t \leq  t5})^2+(\text{log}H_{t5\leq t \leq t10}-\text{log}
L_{t5\leq t \leq  t10})^2+... \\ \nonumber
&&RR_{5,1,1}=(\text{log} H_{t1\leq t \leq t6}-\text{log} L_{t1\leq t \leq  t6})^2+(\text{log}H_{t6\leq t \leq t11}-\text{log}
L_{t6\leq t \leq  t11})^2+... \\ \nonumber
&&RR_{5,1}=\frac{\sum_{i=0}^{4}RR_{5,1,i}} {4 \text{log} (2)5} \nonumber
\end{eqnarray}

{\centering
\section{\normalsize EXPECTILE AND CARE TYPE MODELS}\label{expectile_care_section}
\par
}
\subsection{Expectile}
The $\tau$ level expectile $\mu_{\tau}$, defined by Aigner, Amemiya and Poirier (1976), can be estimated through minimizing the
following expectation:
\begin{equation*}
\text{E} \left[|\tau-I(Y<\mu_{\tau})|(Y-\mu_{\tau})^2 \right] \, ,
\end{equation*}
where $Y$ is a continuous r.v., $\tau \in [0,1]$, $I(Y<\mu_{\tau})$ equals 1 when $Y<\mu_{\tau}$ and $0$ otherwise.
If $Y={y_1, y_2,...y_n}$, the following asymmetric sum of squares equation is employed for $\mu_{\tau}$ in Taylor (2008):
\begin{equation}\label{als_equation}
\sum_{t=1}^{n} |\tau-I(y_t<\mu_{\tau})|(y_t-\mu_{\tau})^2  \, ,
\end{equation}
so that minimizing this equation results in the Asymmetric Least Squares (ALS) estimator of $\mu_{\tau}$.
No distributional assumption is required to estimate $\mu_{\tau}$ here.

As discussed in Section \ref{introduction_sec}, ES is defined as $\text{ES}_{\alpha}= E(Y|Y<Q_{\alpha})$, which stands for the expected
value of $Y$, conditional on the set of $Y$ that is more extreme than the $\alpha$-level quantile of Y, denoted $Q_{\alpha}$.
Newey and Powell (1987) found a general relationship between the expectile and ES: If $\text{E}(Y)=0$, Taylor (2008) showed this
relationship can be formulated as:
\begin{equation}\label{expectile_es_equation}
\text{ES}_{\alpha}=(1+\frac{\tau}{(1-2\tau)\alpha_{\tau}})\mu_{\tau} \, ,
\end{equation}
where $\mu_{\tau}=Q_{\alpha}$; i.e. $\mu_{\tau}$ occurs at the quantile level $\alpha_{\tau}$ of $Y$. Thus, $\mu_{\tau}$ can be used
to estimate the $\alpha$ level quantile $Q_{\alpha}$, and then scaled to estimate the associated ES.

\subsection{CARE type models and Re-GARCH}
Taylor (2008) proposed the CARE type models that have the similar form as the CAViaR type models (Engle and Manganelli 2004), i.e.
symmetric absolute value (SAV), asymmetric (AS) and indirect GARCH (IG). Here we present only the CARE-SAV model:

\noindent
\textbf{CARE-SAV}:
\begin{align*}
 \mu_{t} = \beta_1 + \beta_2 \mu_{t-1} + \beta_3 |r_{t-1}|
\end{align*}
\noindent
where $r_t$ is the day $t$ return, and $\mu_{t}$ is the $\tau$ level expectile for day $t$, while $\tau$ is removed from the notation for the
reason of brevity. Further, Gerlach and Chen (2015) employed the Range in the CARE framework, simply replacing the lagged return
in the CARE-SAV model by the lagged intra-day Range; their paper found that the Ra-CARE type models demonstrated
superiority compared to the return-based CARE models.

Gerlach, Chen, and Lin (2012) extended the CARE model by adding a return equation and Asymmetric Gaussian (AG) errors, showing
that the resulting maximum likelihood estimator (MLE) was equivalent to the ALS estimator. Gerlach and Chen (2015) extended this
framework to a CARE-X model, allowing any realized measure to drive the CARE equation, though they only considered the Range.
The SAV version of their model can be written:

\noindent
\textbf{CARE-X-SAV}
\begin{eqnarray} \label{range_care_sav_equation}
&&r_t= \mu_t+\varepsilon_t \\ \nonumber
&&\mu_{t} = \beta_1 + \beta_2 \mu_{t-1} + \beta_3 x_{t-1} \\ \nonumber
&&\varepsilon_t {\sim} AG(\tau,0,\sigma)
\end{eqnarray}
\noindent
where $AG$ is the Asymmetric Gaussian distribution and $x_t$ is the realized measure at time $t$.
Both the CARE-SAV and CARE-X-SAV can be estimated by ALS, or by
maximum likelihood (ML) assuming the AG error distribution: these estimators are mathematically equivalent. Thus, the AG
is only employed so as to construct a quasi-likelihood function, that has its' mode exactly coinciding with the ALS estimator,
and that subsequently allows a Bayesian estimator, as developed in Gerlach, Chen, and Lin (2012) for CARE models and Gerlach and
Chen (2015) for CARE-X models.

These CARE-type models can all produce one-step-ahead forecasts of $\mu_t$ (expectiles), which can be directly employed as VaR estimates,
by an appropriate choice of $\tau$; more on this later. Then, Equation \ref{expectile_es_equation} can be employed to scale these
expectile forecasts to produce forecasts of ES. This paper extends the CARE-X model class to incorporate a measurement equation, analogous
to the Realized GARCH class of models.

The innovative Realized-GARCH framework was developed in Hansen \emph{et al.} (2011). Comparing to the conventional GARCH model, Re-GARCH
employs a measurement equation, which captures the contemporaneous connection between unobserved volatility and a realized measure
of variance. The superiority of Re-GARCH compared to GARCH has been demonstrated by several authors, including Hansen \emph{et al.} (2011),
Watanabe (2012) and Gerlach and Chao (2016).

\noindent
\textbf{Re-GARCH}
\begin{eqnarray}\label{rgarch}
&& r_t= \sqrt{h_t} z_t \, , \\ \nonumber
&& h_t= \omega +\beta h_{t-1}+ \gamma x_{t-1} \, , \\ \nonumber
&& x_t = \xi +\varphi h_{t}+ \tau_1 z_t + \tau_2 (z_t^2-1)+  \sigma_{\varepsilon} \varepsilon_t \, , \\ \nonumber
\end{eqnarray}
where the 3rd equation is the measurement equation. Here $z_t \stackrel{\rm i.i.d.} {\sim} D_1(0,1)$ and
$ \varepsilon_t \stackrel{\rm i.i.d.} {\sim} D_2(0,1)$; Hansen  \emph{et al.} (2011) made several suggestions, including
$x_t=RV_t$ and focused on $D_1(0,1) = D_2(0,1) \equiv N(0,1)$. Watanabe (2012) further extended the model through
incorporating the Student-t or skewed-t (Hansen, 1994) for $D_1$, also employed in Gerlach and Chao (2016).

The advantage of a measurement equation is that more information about the latent volatility can be incorporated into the likelihood.
Further, asymmetric effects of positive and negative return shocks on volatility are incorporated in an innovative manner. The proposed
Realized CARE model class, which adds both these features to the existing CARE-X framework, is now presented.

\vspace{0.5cm}
{\centering
\section{\normalsize MODEL PROPOSED} \label{model_section}
}
\noindent
Consider a zero-mean return process with conditional distribution $D_1(0,1)$ and conditional volatility at time t given by $h_t$.
The $\alpha$-level quantile is then given by $Q_{\alpha} = \sqrt{h_t} D_1(\alpha)^{-1}$, which in an expectile framework
is also $\mu_t$. Thus, the dynamics on $h_t$ will imply the dynamics on the expectile series $\mu_t$ in a CARE model. Further,
the relationship between the realized measure $x_t$ and $h_t$ also implies a relationship between $x_t$ and $\mu_t$. We thus propose
two Realized CARE models implied by two sets of volatility dynamics.

Under the volatility dynamics in (\ref{rgarch}), the indirect GARCH Realized-CARE-IG is proposed, as follows:

\noindent
\textbf{Realized-CARE-IG (Re-CARE-IG)}
\begin{align}\label{rcareig}
&r_t= \mu_t+\varepsilon_t \\ \nonumber
&\mu_{t}= -\sqrt{\beta_1 + \beta_2 \mu_{t-1}^{2} + \beta_3 x_{t-1}^{2}} \\ \nonumber
&x_t^2= \xi+\phi \mu_t^{2} + \tau_1 \epsilon_t + \tau_2 (\epsilon_t^2-E(\epsilon^2)) + u_t\\ \nonumber
&\varepsilon_t \stackrel{\rm i.i.d.} {\sim} AG(\tau,0,\sigma), u_t \stackrel{\rm i.i.d.} {\sim} D(0,\sigma_{u}^2), \epsilon_t= \frac{r_t} {\mu_t} \, ,
\end{align}
where $r_t= [\text{log}(C_t)-\text{log}(C_{t-1})]\times100$ is the
percentage log-return for day $t$, $x_t$ is the realized measure at time $t$ and it is sufficient for positivity under the square root to
enforce $\beta1>0, \beta2>0, \beta3>0$. In this paper we consider $x_t= Ra_t, RaO_t, \sqrt{RV_t}, \sqrt{RR_t}$ as well as square roots of
the scaled and sub-sampled versions of $RV_t$ and $RR_t$. Also, we make the standard choice $D \equiv N(0,1)$ for the measurement error.
The top three equations in the Realized-CARE above are named as: the \emph{return equation}, the \emph{CARE equation} and the
\emph{measurement equation}, respectively. The measurement equation here captures the contemporaneous dependence between the
expectile $\mu_t$ and realized measure $x_t$, analogous to capturing that between unobserved volatility and the realized measure
in the Re-GARCH framework.

%\bar{\epsilon^2}

If the volatility dynamics are instead those of standard deviation GARCH, the implied Realized-CARE-SAV model is:

\noindent
\textbf{Realized-CARE-SAV (Re-CARE-SAV)}
\begin{eqnarray}\label{rcaresav}
&&r_t= \mu_t+\varepsilon_t \\ \nonumber
&&\mu_{t}= \beta_1 + \beta_2 \mu_{t-1} + \beta_3 x_{t-1}\\ \nonumber
&&x_t= \xi+\phi |\mu_t|+ \tau_1 \epsilon_t + \tau_2 (\epsilon_t^2-E(\epsilon^2)) + u_t \, , \\ \nonumber
\end{eqnarray}
where $\varepsilon_t \stackrel{\rm i.i.d.} {\sim} AG(\tau,0,\sigma)$, $u_t \stackrel{\rm i.i.d.} {\sim} D(0,\sigma_{u}^2)$ and
the choice $D \equiv N(0,1)$ is made here.

There are two return-related "error" series in these Re-CARE models: one is the additive $\varepsilon_t = r_t - \mu_t$,
which is assumed to follow
an AG distribution, so that the MLE coincides with the ALS (in CARE models); The second is the multiplicative
$\epsilon_t= \frac{r_t} {\mu_t}$, that appears in the measurement equation and is employed to capture the well known leverage effect.
Again, if $\mu_t$ is a multiple of $\sqrt{h_t}$ then, we will have $E(\epsilon_t)=0$, as usual, but to keep a
zero mean asymmetry term $(\epsilon_t^2-E(\epsilon^2))$, we need to know $$ E(\epsilon^2) = E\left(\frac{r_t^2}{\mu_t^2}\right) $$.
The Re-CARE model says nothing about this second moment. Thus, we instead substitute an empirical estimate
$E(\epsilon^2) \approx \bar{\epsilon^2}$, being the sample mean of the squared multiplicative errors. We note
that $E(\epsilon_t^2-\bar{\epsilon^2})= 0$ is preserved if $\bar{\epsilon^2}$ is an unbiased estimate. The term
$\tau_1 \epsilon_t + \tau_2 (\epsilon_t^2-\bar{\epsilon^2})$ thus still generates an asymmetric response in volatility to return shocks.
Further, the sign of $\tau_1$ is expected to be opposite that from an Re-GARCH model, since the expectile $\mu_t$ is negative for
the low quantile levels , e.g. $\alpha=1\%$, considered in the paper.

%Through choosing $x_t$ as $Ra_t, \sqrt{RV_t} \; \text{and} \; \sqrt{RR_t}$ respectively, we propose the
%Realized-CARE-SAV-Range (Re-CARE-SAV-Ra),
%Realized-CARE-SAV-Realized Variance (Re-CARE-SAV-RV) and Realized-CARE-SAV-Realized Range (Re-CARE-SAV-RR) models.

The Re-CARE framework can be easily extended into other nonlinear CARE versions, e.g. by choosing the expectile dynamics
in Gerlach, Chen and Lin (2012); however we focus solely on the Re-CARE-SAV type models in this paper.

In order to guarantee that the series $\mu_t$ does not diverge, a necessary condition for both Re-CARE type models is
$\beta_2+\beta_3\phi < 1$, which is subsequently enforced during estimation. This condition can be derived through
substituting the measurement equation into the CARE equation in either (\ref{rcareig}) or (\ref{rcaresav}). The CARE
equation in Re-CARE framework can produce one-step-ahead expectile forecasts (VaR), which can be mapped to
ES forecasts directly through employing Equation (\ref{expectile_es_equation}).

%In this paper, the Realized-GARCH (Hansen \emph{et.al}, 2011) is also adapted by setting the volatility equation as an absolute value
%GARCH specification (Taylor (1986); Schwert (1989)), as follows:
%
%\noindent
%\textbf{Realized-GARCH-Abs (Re-GARCH-Abs)}
%\begin{align*}\label{r_garch_simu}
%&r_t= \sqrt{h_t} \varepsilon_t^{*} \\  \nonumber
%&\sqrt{h_t}= \beta_1^{*}+ \beta_2^{*}\sqrt{h_{t-1}} + \beta_3^{*} x_{t-1}\\  \nonumber
%&x_t= \xi^{*}+\phi^{*} \sqrt{h_t}+ \tau_1 z_t + \tau_2 (z_t^2-1)+ u_t^{*} \\  \nonumber
%&\varepsilon_t^{*} \stackrel{\rm i.i.d.} {\sim} D_1(0,1), u_t^{*} \stackrel{\rm i.i.d.} {\sim} D_2(0,\sigma_{u^{*}}^2)\\  \nonumber
%&x_t= R_t, \sqrt{RV_t}, \sqrt{RR_t},\\
%\end{align*}
%This model allows us to simulate from the Re-CARE-SAV model for the purpose of comparing the likelihood and Bayesian estimators
%for that model.

{\centering
\section{\normalsize LIKELIHOOD AND BAYESIAN ESTIMATION} \label{beyesian_estimation_section}
\par
}
\noindent
\subsection{CARE Likelihood Function with AG}\label{care_likelihood_section}

With $\mathbf{r}= (r_1, r_2, ..., r_n)'$, the ALS as specified in Equation (\ref{als_r_equation}) is employed by Taylor (2008) to
estimate $\mu_{t}$, after the expectile level $\tau$ is estimated through a grid search: $\tau$ is chosen to make the in-sample
violation rate ($\text{VRate} = \frac{1}{n}\sum_{t=1}^n I(r_t<\mu_t(\hat{\beta}))$) as close as possible to the quantile level $\alpha$.
\begin{equation}\label{als_r_equation}
\sum_{t=1}^{n}(|\tau-I(r_t<\mu_t)|(r_t-\mu_t)^2)
\end{equation}
For each grid value of $\tau$, the ALS estimator of the CARE equation parameters $\beta$ is found, yielding an associated VRate($\tau$).
$\hat{\tau}$ is set to the grid value of $\tau$ s.t. VRate is closest to the desired $\alpha$. Then, the ALS estimator of $\hat{\beta}$
conditional on $\hat{tau}$ is found.

Gerlach, Chen and Lin (2012), Gerlach and Chen (2016) develop an asymmetric Gaussian (AG) distribution and include it as the error
distribution in an observation equation for a CARE model, i.e. $\varepsilon_t {\sim} AG(\tau,0,\sigma)$ in (\ref{range_care_sav_equation}).
This makes the construction of a likelihood function feasible. The scale factor $\sigma$ is a nuisance parameter and can be integrated out,
Gerlach, Chen and Lin (2012) employ a Jeffreys prior in this integration and show that maximizing the resulting integrated
likelihood function produces identical estimation as the ALS approach. However, the likelihood formulation also allows access
to powerful computational Bayesian approaches, such as adaptive MCMC algorithms, for estimation.

The CARE (integrated) likelihood in this setting is:
\begin{eqnarray}\label{care_like_equation}
L(\mathbf{r};
\mathbf{\theta})=\big(\sum_{t=1}^{n}|\tau-I(r_t<\mu_{t}(\mathbf{\beta})|(r_t-\mu_{t}(\mathbf{\beta}))^2\big)^{-n/2}
\end{eqnarray}

\subsection{Realized CARE Log Likelihood}

Because the Re-CARE framework has a measurement equation, with $u_t \stackrel{\rm i.i.d.} {\sim} N(0,\sigma_{u}^2)$, the full log-likelihood
function for Re-CARE (as in Model \ref{rcaresav}) is the sum of the log-likelihood $\ell (\mathbf{r};\mathbf{\theta})$ for the
CARE equation and the log-likelihood $\ell (\mathbf{x}|\mathbf{r};\mathbf{\theta})$ from the measurement equation. In the
Re-GARCH framework, the measurement equation variable contributes to volatility estimation, thus the GARCH equation in-sample and
predictive log-likelihood values are improved compared to the traditional GARCH. Thus, we expect the measurement equation in the
Re-CARE to also facilitate an improved estimate $\tau$ and of $\mu_{t}$, leading to more accurate VaR and ES forecasts.

\begin{align*}
&\ell(\mathbf{r},\mathbf{x};\mathbf{\theta})= \ell(\mathbf{r};\mathbf{\theta})+ \ell(\mathbf{x}|\mathbf{r};\mathbf{\theta})\\
&= \underbrace{(-n/2)log\big(\sum_{t=1}^{n}|\tau-I(r_t<\mu_{t}(\mathbf{\beta})|
(r_t-\mu_{t}(\mathbf{\beta}))^2\big)}_{\ell (\mathbf{r};\mathbf{\theta})}\\
& \underbrace{-\frac {1}{2} \sum_{t=1}^{n} \big( log(2 \pi)+log(\sigma_{u}^2)+
   u_t^2/\sigma_{u}^2 \big)}_{\ell (\mathbf{x}|\mathbf{r};\mathbf{\theta})}\\
\end{align*}
where ${\bf u}$ is the measurement equation residual series, e.g. in the Re-CARE-SAV model (\ref{rcaresav}),
$u_t= x_t- \xi- \phi |\mu_t| - \tau_1 \epsilon_t - \tau_2 (\epsilon_t^2-\bar{\epsilon^2})$, $t=1,\ldots,n$.

For the Re-GARCH model framework, Hansen \emph{et.al} (2011) studied the asymptotic properties of the quasi-maximum likelihood estimator,
conjecturing a central limit theorem. Yao and Tong (1996) considered the asymptotics of ALS estimation for expectile regression and showed
consistency of the estimator. Results from both these papers allow us to conjecture the consistency and asymptotic normality of the
ML estimator obtained by numerically maximizing the log-likelihood function above. We leave the proofs for future work.
However, convergence issues in the numerical likelihood optimization, to be discussed, lead us to instead consider MCMC estimation.

\subsection{Bayesian Estimation}
Given a likelihood function, and the specification of a prior distribution, Bayesian algorithms can be employed to estimate the
parameters of an Re-CARE model. A two-step adaptive Bayesian MCMC method, extended from that in Gerlach and Wang (2016) is employed.
First, the parameters are dived into two blocks: $\utwi{\theta_1}=(\beta_1,\beta_2,\beta_3, \phi)^{'}$ and
$\utwi{\theta_2}=(\xi, \tau_1, \tau_2, \sigma)^{'}$, where groupings are
chosen to maximize within group correlation of MCMC iterates; e.g. here the stationarity constraint $\beta_2+\beta_3\phi<1$ induces
some correlation among the three parameters, whilst in GARCH models the equivalent of $\beta_1, \beta_2$ are known to be highly
negatively correlated.

Priors are chosen to be uninformative over the possible stationarity (and positivity, where relevant) regions,
e.g. $\pi(\utwi{\theta})\propto I(A)$, which is a flat prior for $\utwi{\theta}$ over the region $A$.

An adaptive MCMC algorithm, extended from that in Gerlach and Wang (2016), based on that in Chen and So (2006), employs a
random walk Metropolis (RW-M) for the burn-in period and and independent kernel Metropolis-Hastings (IK-MH) algorithm
(Metropolis et al., 1953; Hastings, 1970) for the sampling period.
The burn-in period uses a Gaussian proposal distribution for the random walk process of each parameter group. The covariance matrix
of the proposal distribution in each block is tuned towards a target accept ratio of $23.4\%$ (Roberts, Gelman and Gilks, 1997). Then the
IK-MH sampling period incorporates a mixture of three Gaussian proposal distributions. The sample mean of the last 10\% of the
burn-in period samples are used as the proposal mean vector, while the sample variance-covariance matrix $\Sigma$ is employed so
that the three Gaussian proposal var-cov matrices are: $\Sigma$, $10\Sigma$, $100\Sigma$ respectively, where $\Sigma$ is
calculated as the covariance of the last 10\% of the burn-in period samples, for each block.

\subsection{Expectile Level Search}\label{expectile_level_search}

As discussed in Section \ref{care_likelihood_section}, the estimation of CARE type models relies on a full grid search of the optimal
expectile level $\tau$, e.g employing $M$ equally spaced trial values of $\tau$ on $[0,\alpha]$, as proposed in Taylor (2008).
For each grid value of $\tau$, the ALS estimator of the CARE equation parameters $\beta$ is found, yielding an associated VRate($\tau$).
$\hat{tau}$ is set to the grid value of $\tau$ s.t. VRate is closest to the desired $\alpha$.

This paper extends that approach in three ways. First, the quantile loss function, Equation (\ref{q_loss}), is proposed as a more
suitable optimization criterion than the VRate, which e.g. may not detect autocorrelation in violations.
Further, since quantiles are elicitable, in the sense defined by Gneiting (2011), and the standard quantile loss function is
strictly consistent, i.e. the expected loss is a minimum at the true quantile series, minimizing (\ref{q_loss}) is a more reasonable choice
to estimate $\tau$. Thus, the selected expectile level $\tau$ during the grid search should be the one that minimizes the quantile
loss function:
\begin{equation}\label{q_loss}
\sum_{t=1}^{m}(\alpha-I(y_t<\mu_t))(y_t-\mu_t)  \, ,
\end{equation}
where $\mu_{1},\ldots,\mu_{m}$ is a series of expectiles at level $\alpha$ for the return observations $y_{1},\ldots,y_{m}$.

\begin{figure}[htp]
     \centering
\includegraphics[width=.9\textwidth]{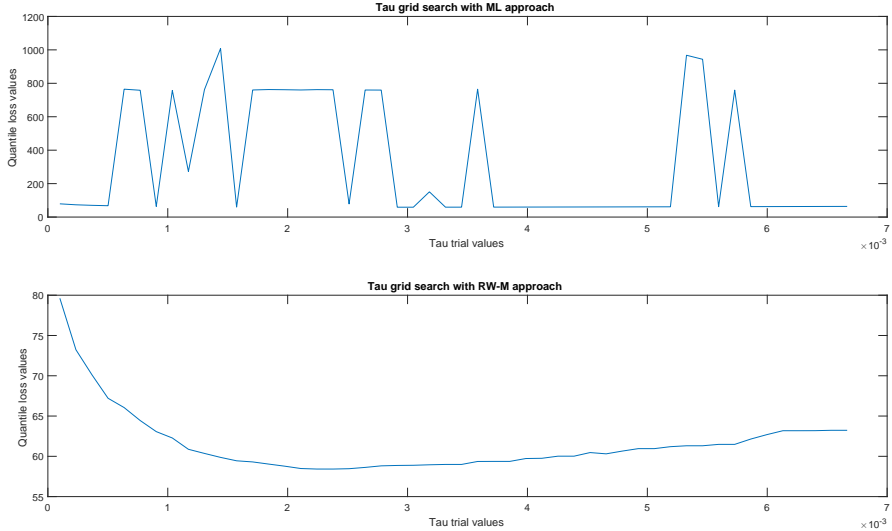}
\caption{\label{Fig_vloss} Re-CARE-RR $\tau$ full grid search with ML and RW-M approaches employ quantile loss function as objective function.}
\end{figure}

Taylor (2008) and Gerlach, Chen, and Lin (2012) both employed ALS during the expectile level grid search procedure. However, employing
the same approach for the Re-CARE model generates abnormal loss function values and tends to suffer from convergence issues.
As an illustration, consider the top plot in Figure \ref{Fig_vloss}, showing the minimized quantile loss function values against the grid
of values for $\tau$ in the Re-CARE-SAV model, for a simulated data set. Though it is possible that a satisfactory estimate of $\tau$
is obtained, it is clear that an un-expectantly non-smooth function over $\tau$ results. Our investigations showed that this non-smoothness,
which occurred in all simulated and real datsets that we tried, was a result of intermittent non-convergence of the ALS optimization,
using the optimization toolbox in Matlab. Further, when the optimization was changed to be via
a genetic algorithm, only minimal improvements were made: in fact, the resulting plots were qualitatively the same and are thus not shown, to save
space. Therefore, the usual search procedure and optimization needs to be adjusted for Re-CARE models.

As an alternative, and the second contribution here, we propose to employ a stochastic optimization algorithm, employing a fast RW-M approach,
to allow improved accuracy in estimation of $\tau$ and convergence in estimating $\beta$ for each grid value of $\tau$.
For each value of $\tau$ considered, the RW-M algorithm from the burn-in period of the full MCMC sampling scheme is employed to give
a few thousand iterates of the parameters, and thus of the series $\mu$, each of which is evaluated via the quantile loss function
in (\ref{q_loss}), with the minimum loss over the iterates eventually chosen; then $\hat{\tau}$ the value of $\tau$ associated with the
overall minimum of these minimum loss values over the grid of $\tau$.
To illustrate the improvement under this approach, the bottom plot of Figure \ref{Fig_vloss} shows the relationship between $\tau$ and
the corresponding minimum loss function, which is now quite smooth and regular, and importantly monotonic, with any remaining non-smoothness being due to the
Monte Carlo error inherent in the RW-M stochastic search method. This sort of improvement is found in all data sets, simulated and real, that we tried.
However, our approach is more time consuming since an MCMC run is required for each grid value of $\tau$.

The third contribution is to make the proposed method faster than a pure grid search, without sacrificing accuracy. To assist in speeding
things up, we propose to find a smaller and more refined region than $[0,\alpha]$ on which to do a more targeted grid-search, via the
following two-step approach. First, a coarse grid search in undertaken and the results employed to identify a smaller area where the
loss function is comparatively low, followed by a refined grid search inside that area.

\noindent
\textbf{Proposed two-step target $\tau$ search approach}\\
\textbf{Step 1}:
Choose $M_1$ equally spaced values for $\tau$, generated in $[m_1, m_2]$; we set $M_1=7$, $m_1=0.0001$, $m_2=\alpha /1.5$, the latter chosen
because the empirical study with full grid search shows $\hat{\tau}$ is well always inside that region.
The minimum loss function (\ref{q_loss}) is calculated for these $M$ values, based on the RW-M search algorithm. For the first
$\tau$ trial value $m_1$, the RW-M is run for a minimum of 10,000 iterations, and a maximum 15,000 iterations, and is stopped if the
maximum of the likelihood function has not changed for more than 1000 iterations; typically this has occurred before the 10000th iterate.
For the remaining $M_1-1$ values, say $\tau_i$, ($i=2,\ldots, M_1$) the estimated MLE of the parameters from $\tau_{i-1}$ are used as the
MCMC starting values, and the RW-M is run for a minimum of 2,000 iterations, with a maximum of 10,000 iterations, and again stopped
when the maximum of the likelihood has not changed for 1000 iterations; this typically happens between 2000 and 4000 iterations.

\begin{figure}[htp]
     \centering
\includegraphics[width=.9\textwidth]{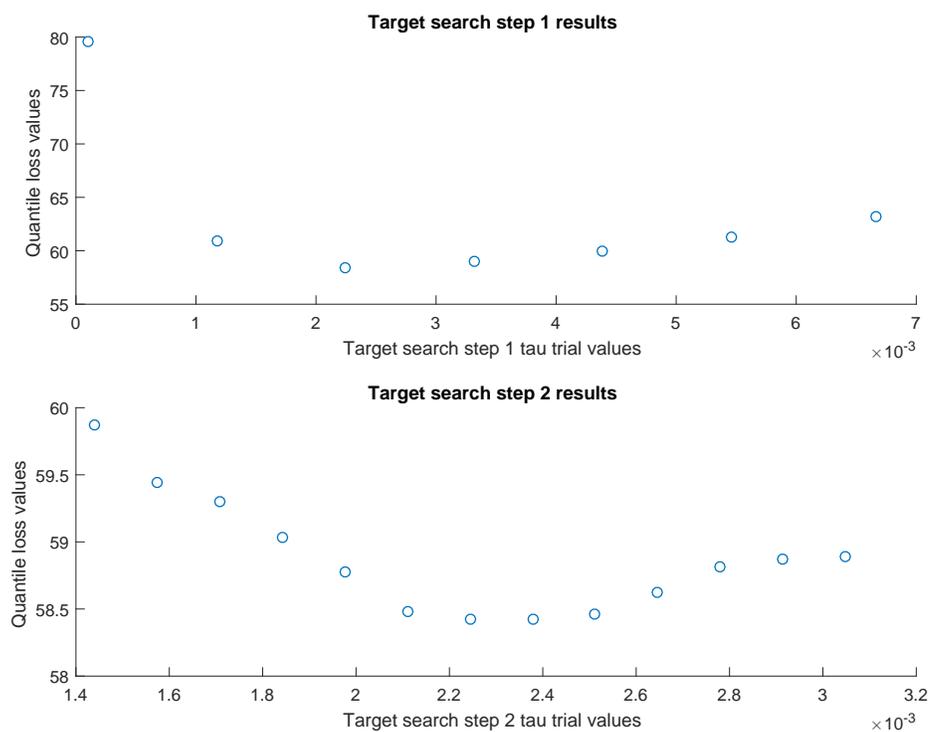}
\caption{\label{Fig_target_search} 2 step $\tau$ target search with RW-M approach employ quantile loss function as objective function.}
\end{figure}

An example of the $M_1=7$ calculated loss function vs $\tau$ values from step 1 is presented in the top plot of
Figure \ref{Fig_target_search}, where $\tau= 0.0022$ generates the minimum loss function value and is selected.

\noindent
\textbf{Step 2}:
A focused, refined grid search is conducted between the two grid values of $\tau$ that are immediately below and above that of the minimum
$\tau$ from step 1. We choose $M_2$ equally spaced points (we set $M_2=6$), with half on either side of the optimum step 1 $\tau$. Then,
the final $\hat{\tau}$ is selected as that value whose associated MLE and series ${\bf \mu}$ minimize the loss function in this
second grid search. An example is shown in the bottom plot of Figure \ref{Fig_target_search}.
The RW-M method is run for the same number of iterations as in Step 1 and with the same settings.

A simulation study will be conducted in Section \ref{simulation_section} in order to study the validity of the proposed 2 step targeted
grid search employing the RW-M method procedure proposed here.

\begin{figure}[htp]
     \centering
\includegraphics[width=.9\textwidth]{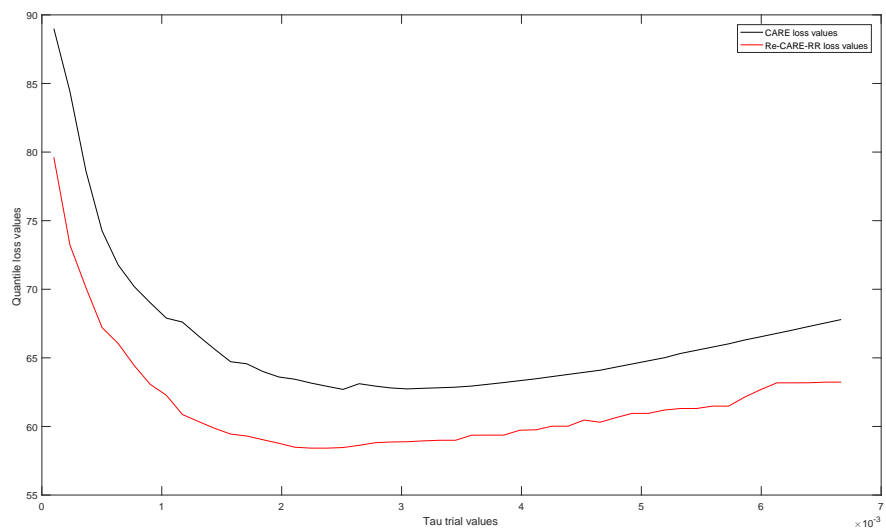}
\caption{\label{Fig_vloss_care} $\tau$ full grid search comparison with CARE ML and Re-CARE-RR RW-M approaches employ quantile loss function as objective function.}
\end{figure}

Finally, employing the quantile loss function (\ref{q_loss}) as the expectile grid search objective function also enables the
statistical comparison between the in-sample quantile estimation accuracy with different CARE-type models.
Figure \ref{Fig_vloss_care} illustrates that the Re-CARE-RR consistently generates smaller quantile loss function values than the
conventional CARE model for every $\tau$ during the grid search process, using with the same S\&P 500 data set, which provides
evidence on the improved in-sample expectile estimation accuracy through employing the proposed Re-CARE framework. Similar results and plots, not shown,
pertain to the other data sets in our empirical study.

{\centering
\section{\normalsize SIMULATION STUDY}\label{simulation_section}
\par
}
\noindent
A simulation study is conducted to compare the properties and performance of the RW-M stochastic search method and maximum
likelihood (equivalent to ALS) estimation approaches for the Re-CARE model, with respect to parameter estimation and
one-step-ahead VaR and ES forecasting accuracy. Both the mean and
Root Mean Square Error (RMSE) values are calculated for the RW-M and ML methods, over the replicated data sets, to illustrate their respective
bias and precision. $N=1000$ simulated datasets are generated from a square root Realized-GARCH model, specified as Model (\ref{r_garch_simu}).
The equivalent Re-CARE-SAV model was fit to each data set, once using ML and once using RW-M. Sample size $n=3000$ is employed for each
simulated data set.

Data replications are simulated from:
\begin{eqnarray}\label{r_garch_simu}
&&r_t= \sqrt{h_t} \varepsilon_t^{*} \\  \nonumber
&&\sqrt{h_t}= 0.02+ 0.75  \sqrt{h_{t-1}} + 0.25 x_{t-1}\\  \nonumber
&&x_t= 0.1+0.9 \sqrt{h_t}-0.02 \varepsilon_t^{*}  + 0.02 (\varepsilon_t^{*2}-1) +u_t \\  \nonumber
&&\varepsilon_t^{*} \stackrel{\rm i.i.d.} {\sim} N(0,1), u_t \stackrel{\rm i.i.d.} {\sim} N(0,0.3^2)\\  \nonumber
\end{eqnarray}

In order to calculate the corresponding Re-CARE-SAV true parameter values, a mapping between from the square root Realized-GARCH to the
Realized-CARE-SAV is required. With $\text{VaR}_t=\mu_t=\sqrt{h_t} \Phi^{-1}(\alpha)$, then
$\sqrt{h_t} =\frac{\mu_t } {\Phi^{-1}(\alpha)}= \frac{\text{VaR}_t} {\Phi^{-1}(\alpha)}$, where $\Phi^{-1}(\alpha)$ is the standard
Normal inverse cdf at $\alpha$ quantile level. Further, with $\varepsilon_t^{*} \stackrel{\rm i.i.d.} {\sim} N(0,1)$, we have
$\epsilon_t= \frac{r_t} {\mu_t} = \frac{r_t} {\sqrt{h_t} \Phi^{-1}(\alpha)} = \frac{\varepsilon_t^{*}} {\Phi^{-1}(\alpha)}$.
Substituting back into the GARCH and measurement equations of Model (\ref{r_garch_simu}),
%results in:
%
%\begin{eqnarray}
%&& \frac{ \mu_{t}} {\Phi^{-1}(\alpha)}= 0.02 + 0.75 \frac{\mu_{t-1}} {\Phi^{-1}(\alpha)}+ 0.25 x_{t-1} \\  \nonumber
%&& x_t= 0.1+0.9 \frac{ \mu_{t}} {\Phi^{-1}(\alpha)}+ u_t \\  \nonumber
%\end{eqnarray}
the corresponding Realized-CARE-SAV specification can be written:

\begin{eqnarray}
&&\mu_{t}= 0.02 \Phi^{-1}(\alpha) + 0.75 \mu_{t-1}+ 0.25 \Phi^{-1}(\alpha) x_{t-1} \\ \nonumber
&&x_t= 0.1-\frac{0.9} {\Phi^{-1}(\alpha)} |\mu_t| -0.02\Phi^{-1}(\alpha)\epsilon_t + 0.02 \Phi^{-1}(\alpha)^2 (\epsilon_t^2-  \frac{1} {\Phi^{-1}(\alpha)^2} ) + u_t \\  \nonumber
\end{eqnarray}
allowing true parameter values to be calculated or read off. These true values appear in Table \ref{simu_table}.

In each model the true one-step-ahead $\alpha$ level VaR forecast is then $\text{VaR}_{n+1}= \sigma_{n+1}\Phi^{-1}(\alpha)$,
and the true one-step-ahead $\alpha$ level ES forecast is $\text{ES}_{n+1}= \sigma_{n+1}\Phi^{-1}(\delta_\alpha)$, where
$\delta_\alpha$ is the quantile level that ES occurs at for the standard normal distribution (Gerlach and Chen, 2016).
Following Basel II and Basel III risk management guidelines, the $1\%$ quantile level is employed
(corresponding $\delta_\alpha=0.38 \%$ with the standard normal distribution), then the true value of
$\text{VaR}_{n+1}$ and $\text{ES}_{n+1}$ can be calculated for each dataset; the averages of these, over
the 1000 datasets, are given in the "True" column of Table \ref{simu_table}. Through the one-to-one relationship
between VaR and ES (Equation (\ref{expectile_es_equation})), the true value of $\tau$ is $0.001452$ for this model.
In addition, the targeted grid search of $\tau$ as presented in Section \ref{expectile_level_search} is incorporated in the RW-M process, while
there is no target search for $\tau$ during the ML estimation, to testify the accuracy of target search.

The Re-CARE-SAV model is fit to the 1000 datasets generated, once using the RW-M method and once using the ML estimator
(the `fmincon' constrained optimisation routine in Matlab software is employed). The RW-M iterations are
specified in Section \ref{expectile_level_search}.

Estimation results are summarized in Table \ref{simu_table}, where boxes indicate the preferred model in terms of minimum
bias (Mean) and maximum precision (minimum RMSE). The results clearly favour the RW-M estimator compared to the MLE; as expected in light of
the convergence issues discussed and illustrated in Section \ref{expectile_level_search}. The bias results favor the MCMC approach in 9 out of 9
parameter estimates and VaR\& ES forecasts. Further, the precision is clearly higher for the MCMC method for all 9 parameters and both tail risk forecasts, while the RMSE of $\tau$ from
RW-M is only marginally higher than that from ML. %In other words, a very small price in accuracy of $\tau$ is paid to get vastly
%favourable parameter estimation results.
%It is worth note that ML approach generated relatively accurate $\tau$ search results in the simulation study mainly because the true
%expectile level is the same (0.001452) for every simulated data set. However, in the empirical study, the estimated $\tau$ will be changed with
%different forecasting steps, the convergence issue identified in Figure \ref{Fig_vloss} will generate very inaccurate $\tau$
%estimation results, leading to inaccurate tail risk forecasts.
Finally, the estimation results for $\tau$ with RW-M stochastic search approach highlight the validity of
the proposed targeted search approach. The targeted procedure was only used for the RW-M expectile search process. After $\tau$ was
selected with target search, the adaptive Bayesian approach as described in Section \ref{beyesian_estimation_section} is employed for the
parameters estimation. %The MCMC sampler has 10000 iterations for each data set, with a burn-in of 5000 iterations.

\begin{table}[!ht]
\begin{center}
\caption{\label{simu_table} \small Summary statistics for the two estimators of the Realized-CARE-SAV model, with data simulated from Model \ref{r_garch_simu}.}\tabcolsep=10pt

\begin{tabular}{lcccccccc} \hline
$n=3000$               &             & \multicolumn{2}{c}{RW-M}      &  \multicolumn{2}{c}{ML}   \\
Parameter              &True         &Mean           &  RMSE         &Mean           & RMSE    \\ \hline
$\beta_1$              & -0.0465	 &\fbox{-0.0432} &\fbox{0.1286}	 &-0.2337	     &0.6295 \\
$\beta_2$              & 0.7500	     &\fbox{0.7414}	 &\fbox{0.0280}	 &0.7305	     &0.0800 \\
$\beta_3$              & -0.5816     &\fbox{-0.6021} &\fbox{0.1004}  &-0.5358        &0.3925 \\
$\xi$                  & 0.1000	     &\fbox{0.1056}	 &\fbox{0.1908}	 &0.5490         &1.5085 \\
$\varphi$              & 0.3869	     &\fbox{0.3875}  &\fbox{0.0484}  &0.2884	     &0.3318 \\
$\tau_1$               & 0.0465      &\fbox{0.0463}	 &\fbox{0.0132}  &0.0469         &0.0134 \\
$\tau_2$               & 0.1082	     &\fbox{0.1074}	 &\fbox{0.0220}	 &0.1011         &0.0402 \\
$\sigma_{u}$           & 0.3000      &\fbox{0.3007}  &\fbox{0.0041}  &0.3019         &0.0073 \\
$\tau$                 & 0.001452    &\fbox{0.001474}&\fbox{0.000313}&0.001405       &0.000348 \\
$\text{VaR}_{n+1}$     & -4.1751     &\fbox{-4.1768} &\fbox{0.2513}	 &-4.2617        &0.4970 \\
$\text{ES}_{n+1}$      & -4.7831	 &\fbox{-4.7929} &\fbox{0.2773}  &-4.8600        &0.5601  \\
\hline
\end{tabular}
\end{center}
\emph{Note}:\small  A box indicates the favored estimators, based on mean and RMSE.
\end{table}

{\centering
\section{\normalsize DATA and EMPIRICAL STUDY}\label{data_empirical_section}
\par
}
\subsection{Data Description}
Daily and high frequency data, observed at 1-minute and 5-minute frequency, including daily open, high, low and closing prices, are downloaded from
Thomson Reuters Tick History. Data are collected for 7 market indices: S\&P500, NASDAQ (both US), Hang Seng (Hong Kong), FTSE 100 (UK),
DAX (Germany), SMI (Swiss) and ASX200 (Australia), with time range Jan 2000 to June 2016; as well as for 2 individual assets: IBM and GE (both US).
IBM has the same starting data as 7 indices, while the starting data collection time for GE is May 2000, only after its $3:1$ stock split in April, 2000.

\begin{figure}[htp]
     \centering
\includegraphics[width=.9\textwidth]{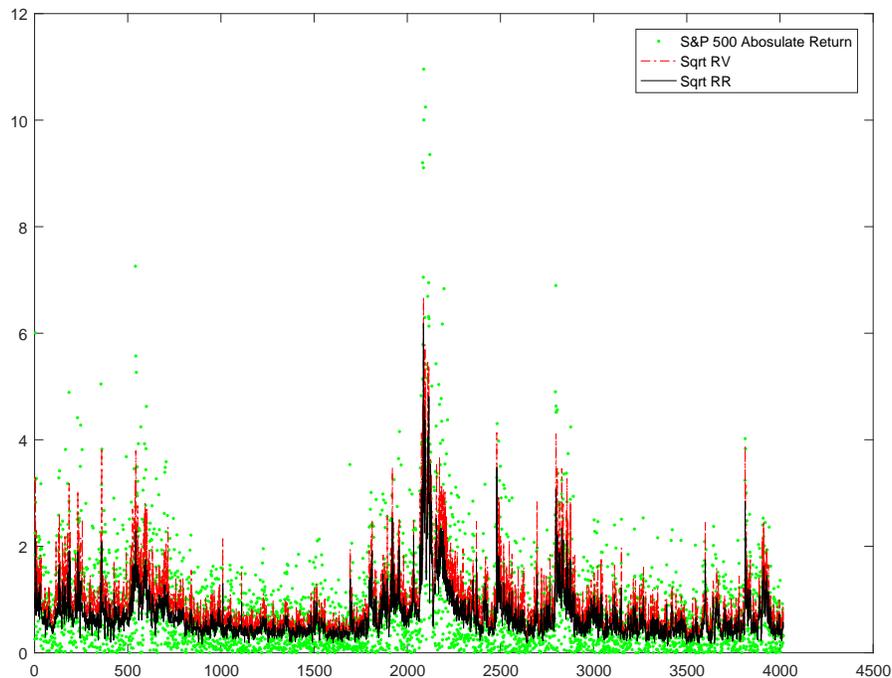}
\caption{\label{Fig2} S\&P 500 Abs Return, Sqrt RV and Sqrt RR Plots.}
\end{figure}

The data are used to calculate the daily return, daily range and daily range plus overnight price jump. Further, the 5-minute data
are employed to calculate the daily RV and RR measures, while both 5 and 1-minute data are employed to produce daily scaled and sub-sampled
versions of these two measures, as in Section \ref{realized_measure_section}; $q=66$ is employed for the scaling process, i.e. around 3 months.
Thus, the final starting time is 3 months from the starting time of data collection. Figure \ref{Fig2} plots the S\&P 500 absolute value of daily
returns, as well as the square root of RV and square root of RR, for exposition.

\subsection{Tail Risk Forecasting}

Both daily Value-at-Risk (VaR) and Expected Shortfall (ES) are estimated for the 7 indices and the 2 asset series, as recommended in
the Basel II and III Capital Accord. As discussed in Section \ref{expectile_care_section}, in the CARE setting the VaR, which is the $\alpha$
level quantile, can be estimated by the corresponding $\tau$ level expectile. Then ES can subsequently be calculated through employing the
one-to-one relationship in equation (\ref{expectile_es_equation}), .

A rolling window with fixed size in-sample data is employed for estimation to produce each 1 step ahead forecast; the in-sample size $n$ is
given in Table \ref{var_fore_table} for each series, which differs due to non-trading days in each market. In order to see the
performance during the GFC period, the initial date of the forecast sample is chosen as the beginning of 2008. On average,
2111 VaR and ES forecasts are generated for each return series from a range of models. These include the proposed Re-CARE type models
(estimated with MCMC) with different input measures of volatility: range, range considering overnight jump, RV \& RR, scaled RV \& RR and
sub-sampled RV \& RR. The conventional GARCH, EGARCH and GJR-GARCH with
Student-t distribution,  CARE-SAV and Re-GARCH with Gaussian and Student-t observation equation error distributions,
are also included, for the purpose of comparison. Further, a filtered GARCH (GARCH-HS) approach is also included,
where a GARCH-t is fit to the in-sample data, then a standardised VaR and ES are estimated via historical simulation (using all the in-sample data)
from the sample of returns (e.g. $r_1,\ldots,r_n$ divided by their GARCH-estimated conditional standard deviation (i.e. $r_t/\sqrt{\hat{h_t}}$).
Then final forecasts of VaR, ES are found by multiplying the standardised VaR, ES estimates by the forecast $\sqrt{\hat{h}_{n+1}}$ from the GARCH-t model.
All these aforementioned models are estimated by ML, using the Econometrics toolbox in Matlab (GARCH-t, EGARCH-t, GJR-t and GARCH-HS) or code
developed by the authors (CARE-SAV and Re-GARCH). The actual forecast sample sizes $m$, in each series, are given in Table \ref{var_fore_table}.

The VaR violation rate (VRate) %and ES violation rate (ESRate) are
is employed to initially assess the VaR forecasting accuracy. VRate is simply the proportion of returns that exceed the forecasted VaR
level in the forecasting period, given in equation (\ref{varvrate_equation})% and (\ref{esvrate_equation}).
Models with VRate closest to nominal quantile level $\alpha=0.01$ are preferred.

%In addition, Gerlach and Chen (2016) presented the quantile levels where the 1\% ES is estimated to fall at over many different
%distributional choices. For the Gaussian this is $0.38\%$. For other distributions estimated in GARCH models for daily return data, the
%$1\%$ ES quantile is estimated in $0.35 \%$ to $0.37\%$, for a range of non-Gaussian distributions. As such, following Gerlach and Chen (2016),
%$0.36\%$ is chosen as the approximate nominal expected ESRate for the ES forecasting study for CARE-type models.

\begin{equation}\label{varvrate_equation}
\text{VRate}= \frac{1}{m} \sum_{t=n+1}^{n+m} I(r_t<\text{VaR}_t)\, ,
\end{equation}
where $n$ is the in-sample size and $m$ is the forecasting sample size.
%\begin{equation}\label{esvrate_equation}
%\text{ESRate}= \frac{1}{m} \sum_{t=n+1}^{n+m} I(r_t<\text{ES}_t),
%\end{equation}

However, having a VRate close to the expected level is a necessary but not sufficient condition to guarantee an accurate forecasting model.
Thus several standard quantile accuracy and independence tests are also employed: e.g. the unconditional coverage (UC) and conditional
coverage (CC) tests of Kupiec (1995) and Christoffersen (1998) respectively, as well as the dynamic quantile (DQ) test of
Engle and Manganelli (2004) and the VQR test of Gaglione et al. (2011). %With the approach of Gerlach and Chen (2016), the derived expected ES level can be used so as to treat
%ES forecasts as quantile forecasts at appropriate quantile levels and these same tests can be applied;
Further, the standard bootstrap t-test that the ES residuals for VaR violations have mean 0 is applied to test each model's ES forecast
series.

\subsubsection{\normalsize Value at Risk}

Table \ref{var_fore_table} presents the VRates at the 1\% quantile for each model over the 9 return series,
while Table \ref{Summ_var_fore} summarizes those results. A box indicates the model that has observed VRate closest to 1\% in
each market, while bolding indicates the model with VRate furthest from 1\%. The G-t, EGARCH-t, GJR-t, CARE-SAV, Re-GARCH-GG and
Re-GARCH-tG with RR are estimated with ML, and the Realized-CARE type models are estimated with MCMC, incorporating the RW-M target
search approach testified in Section \ref{simulation_section}.

Clearly from Table \ref{var_fore_table}, Re-CARE models as a group have most of the optimal VaR forecast series, in terms of being
closest to VRate of 1\%, over the 9 return series. From Table \ref{Summ_var_fore} Re-CARE models employing either RaO and RR
have VRates closest to the 1\% quantile level on average and via the median; though most Re-CARE models were close to 1\% on this measure
and most other models were not. All the models were anti-conservative, having VRates on average (and median) above 1\%: Re-GARCH-GG was
most ant-conservative, generating 80-90\% too many violations, not surprising since it is the only model
employing the 'straw man' Gaussian error distribution.

Chang et al. (2011) and McAleer et al. (2013) proposed using forecast combinations of the VaR series from different models, to take advantage
of associated empirically-observed efficiencies from forecast combination, but also to potentially robustify against the effects of
financial crises like the GFC. This approach is employed here: specifically, the four series created by taking the mean ("FC-Mean"),
median ("FC-Med"), minimum ("FC-Min") and maximum ("FC-Max") of the VaR forecasts from all 15 models for each day, are considered.
The lower tail VaR forecasts are considered here, so "FC-Min" is the most extreme of the 15 forecasts (i.e. furthest from 0) and
"FC-Max" is the least extreme. The VRates for "FC-Mean", "FC-Med", "FC-Min" and "FC-Max" series are also presented in
Tables \ref{var_fore_table} and \ref{Summ_var_fore}. Regarding these,
the "FC-Min" approach is highly conservative in each series, with few if any violations, while the "FC-Max" series produces anti-conservative
VaR forecasts that generate far too many violations. The "FC-Mean" and "FC-Median" of the 15 models produced series that generate
very competitive VRates.

\begin{table}[!ht]
\begin{center}
\caption{\label{var_fore_table} \small 1\% VaR Forecasting VRate with different models on 7 indices and 2 assets.}\tabcolsep=10pt
\tiny
\begin{tabular}{lccccccccccc} \hline
Model       &S\&P 500       &NASDAQ          &HK              &FTSE           &DAX             &SMI            &ASX200          &IBM            &GE    \\ \hline
G-t         &\bf{1.467\%}   &\bf{1.895}\%    &\bf{1.652\%}    &\bf{1.731\%}   &1.362\%         &\bf{1.617\%}   &\bf{1.702\%}    &1.183\%        &\fbox{0.945\%}\\
EG-t        &\bf{1.514\%}   &\bf{1.611\%}    &1.215\%         &\bf{1.777\%}   &1.408\%         &\bf{1.712\%}   &\bf{1.466\%}    &1.183\%        &1.040\%\\
GJR-t       &\bf{1.467\%}   &\bf{1.563\%}    &1.263\%         &\bf{1.777\%}   &1.408\%         &\bf{1.759\%}   &\bf{1.513\%}    &1.088\%        &1.040\%\\
Gt-HS       &1.230\%        &\bf{1.563\%}    &1.263\%         &1.123\%        &1.127\%         &1.284\%        &\fbox{0.898\%}  &1.041\%        &1.181\%\\
CARE        &1.278\%        &\bf{1.563\%}    &1.020\%         &1.310\%        &1.221\%         &1.284\%        &1.229\%         &1.183\%        &1.371\%\\
RG-RV-GG    &\bf{2.130\%}   &\bf{1.942\%}    &\bf{2.818\%}    &\bf{1.777\%}   &\bf{2.300\%}    &\bf{1.807\%}   &\bf{1.560\%}    &1.419\%        &1.323\%\\
RG-RV-tG    &\bf{1.467\%}   &\fbox{1.326\%}  &\bf{1.992\%}    &1.310\%        &\bf{1.596\%}    &1.141\%        &1.229\%         &0.851\%        &0.803\%\\
RC-Ra       &\fbox{1.041\%} &\bf{1.563\%}    &1.020\%         &0.795\%        &1.268\%         &\bf{1.474\%}   &\fbox{0.898\%}  &1.041\%        &1.229\%\\
RC-RaO      &\fbox{1.041\%} &\fbox{1.326\%}  &0.875\%         &0.795\%        &\fbox{1.033\%}  &0.808\%        &0.851\%         &1.230\%        &1.229\%\\
RC-RV       &1.278\%        &\bf{1.563\%}    &\bf{2.041\%}    &0.935\%        &1.268\%         &1.189\%        &\fbox{0.898\%}  &1.183\%        &1.087\%\\
RC-RR       &\fbox{1.041\%} &\bf{1.468\%}    &1.118\%         &0.702\%        &1.221\%         &1.427\%        &0.709\%         &\fbox{0.993\%} &1.087\%\\
RC-ScRV     &1.372\%        &\bf{1.705\%}    &1.118\%         &\fbox{0.982\%} &1.362\%         &\fbox{1.046\%} &0.851\%         &1.277\%        &1.040\%\\
RC-ScRR     &1.278\%        &\bf{1.611\%}    &\fbox{0.923\%}  &0.935\%        &\bf{1.455\%}    &1.284\%        &0.662\%         &\fbox{0.993\%} &1.087\%\\
RC-SubRV    &1.325\%        &\bf{1.611\%}    &1.166\%         &0.889\%        &1.362\%         &1.189\%        &0.851\%         &\fbox{0.993\%} &1.181\%\\
RC-SubRR    &1.088\%        &\bf{ 1.516\%}   &1.020\%         &0.795\%        &1.362\%         &1.379\%        &0.662\%         &\fbox{0.993\%} &1.134\%\\
FC-Mean     &1.183\%        &\bf{1.468\%}    &1.166\%         &1.029\%        &1.127\%         &1.189\%        &0.757\%         &\fbox{0.993\%} &\bf{0.567\%}\\
FC-Med      &1.183\%        &1.421\%         &1.118\%         &1.029\%        &1.268\%         &1.236\%        &0.757\%         &1.088\%        &\fbox{0.945\%}\\
FC-Min      & \bf{0.284\%}  &\bf{0.474\%}    &\bf{0.292\%}    &\bf{0.374\%}   &\bf{0.329\%}    &\bf{0.428\%}   &\bf{0.189\%}    &0.615\%        &\bf{0.331\%}\\
FC-Max      & \bf{3.029\%}  &\bf{3.174\%}    &\bf{3.353\%}    &\bf{3.04\%}    &\bf{3.005\%}    &\bf{2.806\%}   &\bf{3.026\%}    &\bf{1.939\%}   &\bf{2.883\%}\\
m           &2113           &2111            &2058            &2138           &2130            &2103           &2115            &2114           &2116\\
n           &1905           &1892            &1890            &1943           &1936            &1930           &1871            &1916           &1839\\
\hline
\end{tabular}
\end{center}
\emph{Note}:\small  Box indicates the favored models based on VRate, in each series, whilst bold indicates the violation rate is
significantly different to 1\% by the UC test. $m$ is the out-of-sample size, and $n$ is in-sample size. RG stands for the Realized-GARCH
type models, and RC represents the Realized-CARE type models. ‘FC’ stands for forecast combination.
\end{table}

\begin{table}[!ht]
\begin{center}
\caption{\label{Summ_var_fore} \small Summary of 1\% VaR Forecast VRates, for different models on 7 indices and 2 assets.}\tabcolsep=10pt
\begin{tabular}{lccc} \hline
%         &      \multicolumn{2}{c}{VaR}       %   &   \multicolumn{2}{c}{ES}     \\
Model    &Mean              &Median          & RMSE \\ \hline %   &Mean              &Median  \\ \hline
G-t      &1.505\%           &1.608\%         &     0.0058 \\ %   &0.563\%           &0.520\%\\
EG-t     &1.437\%           &1.466\%         &     0.0050 \\ %   &0.521\%           &0.520\%\\
GJR-t    &1.432\%           &1.466\%         &     0.0050 \\ %   &0.553\%           &0.568\%\\
Gt-HS    &1.190\%           &1.183\%         & \cb{0.0026} \\ %   &0.421\%           &0.473\%\\
CARE     &1.274\%           &1.277\%         &     0.0031 \\ %   &0.474\%           &0.473\%\\
RG-RV-GG &\cred{1.895\%}    &\cred{1.798\%}  &\cred{0.010} \\ %   &\cred{0.99\%}     &\cred{0.993\%}\\
RG-RV-tG &1.300\%           &1.325\%         &     0.0046 \\ %   &0.458\%           &0.426\%\\
RC-Ra    &1.147\%           &\fbox{1.041\%}  &     0.0028 \\ %   &0.29\%            &0.284\%\\
RC-RaO   &\fbox{1.021\%}    &\fbox{1.041\%}  & \fbox{0.0019} \\ %   &0.237\%           &0.284\%\\
RC-RV    &1.269\%           &1.183\%         &     0.0043 \\ %   &0.384\%           &\cb{0.378\%}\\
RC-RR    &\cb{1.084\%}      &\cb{1.088\%}    &     0.0027 \\ %   &\cb{0.347\%}      &0.284\%\\
RC-ScRV  &1.195\%           &\cb{1.088\%}    &     0.0031 \\ %   &\fbox{0.353\%}    &\cb{0.378\%}\\
RC-ScRR  &1.137\%           &\cb{1.088\%}    &     0.0031  \\ %   &0.342\%           &\fbox{0.331\%}\\
RC-SubRV &1.174\%           &1.183\%         &     0.0029  \\  %   &\fbox{0.353\%}    &\fbox{0.331\%}\\
RC-SubRR &1.105\%           &\cb{1.088\%}    &     0.0028  \\  %   &0.316\%           &0.284\%\\
FC-Mean  &\cb{1.053\%}      &1.135\%         &     0.0025 \\ %   &0.332\%           &\fbox{0.331\%}\\
FC-Med   &1.116\%           &\cb{1.088\%}    & \cb{0.0022} \\ %   &\cb{0.347\%}      &\cb{0.378\%}\\
FC-Min   &0.368\%           &0.331\%         &     0.0064 \\ %   &0.095\%           &0.047\%\\
FC-Max   &\bf{2.916\%}      &\bf{3.027\%}    & \bf{0.0195} \\ %   &\bf{1.311\%}      &\bf{1.325\%}\\
m        &2110.889          &2114&\\
n        &1902.444          &1905&\\
\hline
\end{tabular}
\end{center}
\emph{Note}:\small  Box indicates the favoured model, blue shading indicates the 2nd ranked model, bold indicates the least favoured model,
red shading indicates the 2nd lowest ranked model, in each column. RMSE employs 1\% as the target VRate.
\end{table}

Several tests are employed to statistically assess the forecast accuracy and independence of violations from each VaR forecast model.
Table \ref{var_backtest_table} shows the number of return series (out of 9) in which each 1\% VaR forecast model is rejected for each test,
conducted at a 5\% significance level. The Re-CARE type models are generally less likely to be rejected by the back tests
compared to other models, and the RG-RV-tG, Re-CARE-SubRR, "FC-Mean" and "FC-Med" achieved the least number of rejections (3).
The G-t, "FC-Min" and "FC-Max" combinations are rejected in all 9 series, the EG-t and Re-GARCH-GG models are rejected in 8 series,
respectively. Further, the Re-CARE-RaO and RC-SubRV are both rejected in 7 series, though they generated close to 1\% VaR VRate on average
and by RMSE.

\begin{table}[!ht]
\begin{center}
\caption{\label{var_backtest_table} \small Counts of 1\% VaR  rejections with UC, CC, DQ and VQR tests for different models on 7 indices and 2 assets.}\tabcolsep=10pt
\footnotesize
\begin{tabular}{lcccccc} \hline
Model    &  UC & CC & DQ & DQ4 & VQR  & Total \\ \hline
G-t      &6&6&7&7&5&\bf{9}\\
EG-t     &5&3&4&7&2&\cred{8}\\
GJR-t    &5&3&6&5&3&7\\
Gt-HS    &1&1&1&3&1&\cb{4}\\
CARE     &1&1&0&5&0&5\\
RG-RV-GG &7&7&7&7&5&\cred{8}\\
RG-RV-tG &3&2&2&1&3&\fbox{3}\\
RC-Ra    &2&2&4&6&3&6\\
RC-RaO   &0&0&5&6&7&7\\
RC-RV    &2&2&3&3&3&\cb{4}\\
RC-RR    &1&2&2&4&1&\cb{4}\\
RC-ScRV  &1&1&4&5&1&6\\
RC-ScRR  &2&2&2&3&0&\cb{4}\\
RC-SubRV &1&2&3&6&2&7\\
RC-SubRR &1&2&2&3&0&\fbox{3}\\
FC-Mean  &2&0&0&2&3&\fbox{3}\\
FC-Med   &0&0&0&2&2&\fbox{3}\\
FC-Min   &8&8&7&4&9&\bf{9}\\
FC-Max   &9&9&9&9&9&\bf{9}\\
\hline
\end{tabular}
\end{center}
\emph{Note}:\small Box indicates the model with least number of rejections, blue shading indicates the model with 2nd least number of rejections, bold indicates the model with the highest number of rejections, red shading indicates the model 2nd highest number of rejections.
All tests are conducted at 5\% significance level.
\end{table}

\subsubsection{\normalsize Expected Shortfall}

One-step-ahead daily ES forecasts are generated for the same 15 models and 9 series during
the forecast sample periods. %Regarding the expected level of ESRate for different models and distributions, Chen, Gerlach and Lu (2012) discuss

\begin{figure}[htp]
     \centering
\includegraphics[width=1.1\textwidth]{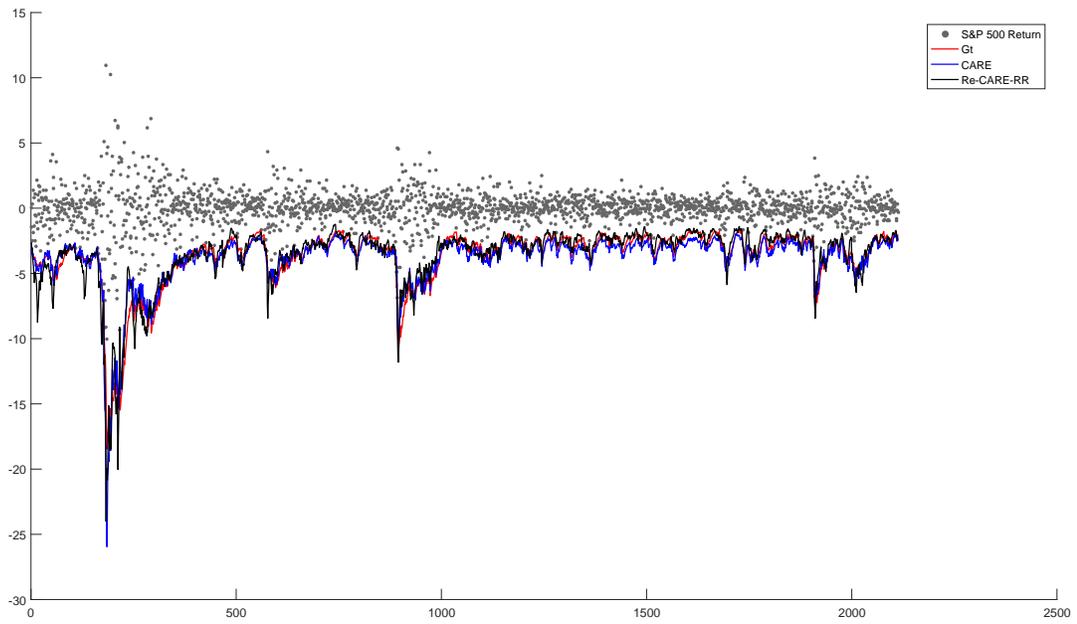}
\caption{\label{Fig_es_fore} S\&P 500 ES Forecasts with Gt, CARE-SAV and Re-CARE-RR.}
\end{figure}

\begin{figure}[htp]
     \centering
\includegraphics[width=1.1\textwidth]{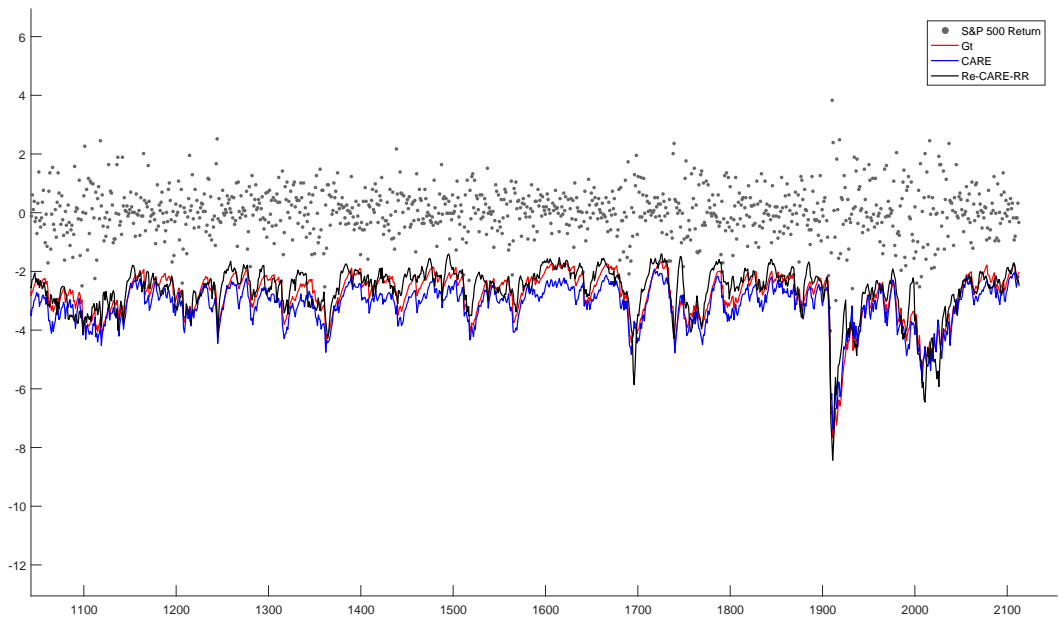}
\caption{\label{Fig_es_fore_zoom_in} S\&P 500 ES Forecasts with Gt, CARE-SAV and Re-CARE-RR (Zoomed in).}
\end{figure}

First, Figure \ref{Fig_es_fore} and \ref{Fig_es_fore_zoom_in} demonstrate the extra efficiency that can be gained by
employing the Re-CARE framework with RR.
Specifically, the ES violation rate of the G-t, CARE-SAV and Re-CARE-RR models are 0.568\%, 0.284\% and 0.379\% respectively
for S\&P500. Employing the typical degrees of freedom estimate for the G-t model, the quantile level that the G-t ES should fall at is
approximately 0.36\%.  %These violation rates mean the G-t generated anti-conservative ES forecasts that produced too many violations, while CARE-SAV is
%more conservative and close to nominal level violation rate and the Re-CARE-RR is slightly anti-conservative here.
Through close inspection of Figure \ref{Fig_es_fore_zoom_in}, e.g. second half of the forecasting period,
the CARE-SAV has an obviously more extreme (in the negative direction) level of ES forecasts than G-t does,
but this also means the capital set aside by financial institutions to cover extreme losses, based on such ES forecasts, is at a higher
level with the CARE-SAV than with G-t, as expected since the CARE-SAV generates fewer violations than the G-t. However, we can clearly
observe the Re-CARE-RR produce ES forecasts that are less extreme than both the CARE-SAV and G-t models here, meaning that lower
amounts of capital are needed to protect against market risk, while simultaneously producing a violation rate much lower than the G-t
(though higher than CARE-SAV) and at the expected rate suggested by the G-t model. This suggests a higher level of efficiency for
the Re-CARE-RR model, at least compared to the G-t, in that this model can produce ES forecasts that have far fewer violations, but are
simultaneously less extreme than those of the traditional GARCH model.
Since the capital set aside by financial institutions should be directly proportional to the ES forecast, the Re-CARE-RR model is
saving these companies money, by giving potentially more accurate and often less extreme ES forecasts; this extra efficiency is also
often observed for the Re-CARE type models in the other markets/assets, especially those employing RR and sub-sampled RR.

Further, during the GFC, when there is a persistence of extreme returns, close inspection of Figure \ref{Fig_es_fore} reveals that the
Re-CARE-RR ES forecasts "recover" the fastest among the 3 models presented, in terms of being marginally the fastest to produce
forecasts that again follow the tail of the data. Traditional GARCH models tend to over-react to extreme events and to be subsequently
very slow to recover, due to their oft-estimated very high level of persistence; Re-CARE models improve on this aspect.

Back testing is conducted on all the ES forecasts using the bootstrap t-test. Based on this test, the worst model is the RC-RaO which
is rejected in 4 of the 9 series. The Gt-HS, CARE, RC-RR, RC-SCRR, RC-SubRV, RC-SubRR were not rejected in any of the series, as was
the forecast combined FC-Med method. The G-t, EG-t, GJR-t, RG-RV-tG, RC-RV, RC-ScRV and FC-Mean models were rejected in only 1 series.
It is clear that this test is not a strong discriminator between these models, but that the Re-CARE models with RV and RR based measures
are mostly not rejected, whilst the previously promising RC-RaO method, in terms of VRates, is again rejected in several more series than
the other Re-CARE models.

\subsubsection{\normalsize VaR\&ES Joint Loss Function}
Cost or loss measures can be applied to assess ES forecasts, as in So and Wong (2011) who employed RMSE and MAD of the
``ES residuals'' $y_t-ES_t$, only
for days when the return violates the associated VaR forecast, i.e. $y_t<VaR_t$. However, these loss functions are not minimized
by the true ES series; in fact Gneiting (2011) showed that ES is not "elicitable": i.e. there is no loss function that is minimized by the
true ES series, in general. Recently, however, Fissler and Ziegel (2016) developed a family of loss functions, that are a joint function of
the associated VaR and ES series. This loss function family are minimized by the true VaR and ES series, i.e. they are strictly
consistent scoring functions for (VaR, ES) considered jointly. The function family form is:
\begin{eqnarray*}
S_t(y_t, VaR_t, ES_t) &=& (I_t -\alpha)G_1(VaR_t) - I_tG_1(y_t) +  G_2(ES_t)\left(ES_t-VaR_t + \frac{I_t}{\alpha}(VaR_t-y_t)\right) \\
                      &-& H(ES_t) + a(y_t) \, ,
\end{eqnarray*}
where $I_t=1$ if $y_t<VaR_t$ and 0 otherwise for $t=1,\ldots,T$, $G_1()$ is increasing, $G_2()$ is strictly increasing and strictly convex,
$G_2 = H^{'}$ and $\lim_{x\to -\infty} G_2(x) = 0$ and $a(\cdot)$ is a real-valued integrable function. Motivated by a suggestion in
Fissler and Ziegel (2016), making the choices: $G_1(x) =x$,
$G_2(x) = exp(x)$, $H(x)= exp(x)$ and  $a(y_t) = 1-\log (1-\alpha)$, which satisfy the required criteria, returns the
scoring function:
\begin{eqnarray}\label{eqveloss}
%\nonumber S_t(y_t, VaR_t, ES_t) &=& \frac{y_t}{ES_t} - \frac{(y_t-VaR_t)(\alpha-I_t)}{\alpha ES_t} - \log \left(\frac{\alpha-1}{ES_t}\right) \\
\nonumber S_t(y_t, VaR_t, ES_t) &=& (I_t -\alpha)VaR_t - I_ty_t  + \exp (ES_t) \left(ES_t-VaR_t + \frac{I_t}{\alpha}(VaR_t-y_t)\right) \\
                                &-& \exp (ES_t) + 1-\log (1-\alpha) \, ,
\end{eqnarray}
where the loss function is $S = \sum_{t-1}^T S_t$. Here, $S$ is a strictly consistent scoring rule that is jointly minimized by the
true VaR and ES series; we use this to informally and jointly assess and compare the VaR and ES forecasts from all models.

Table \ref{veloss} shows the loss function values $S$, calculated using equation (\ref{eqveloss}), which jointly assess the
accuracy of each model's VaR and ES series, during the forecast period for each market. On this measure, the Re-CARE models using RR and
SubRV do best overall, having lower loss than most other models in most series and
being consistently ranked lower on that measure. The EG-t model ranks lowest among individual models, only trailed by the forecast
combination method "FC-Max". Generally the Re-CARE models are higher ranked, having lower loss,
than other models in most markets. These models, together with the "FC-Med" and "FC-Mean", consistently outperform the
all other models.

\begin{table}[!ht]
\begin{center}
\caption{\label{veloss} \small VaR and ES joint loss function values, using equation (\ref{eqveloss}), across the markets; $\alpha=0.01$.}\tabcolsep=10pt
\tiny
\begin{tabular}{lccccccccccccc} \hline
Model     &S\&P 500         &NASDAQ         &HK             &FTSE           &DAX            &SMI            &ASX200         &IBM            &GE    \\ \hline
G-t       &2119.2           &2157.1         &2135.8         &2156.4         &2226.7         &2153.9         &2082.4         &2270.9         &2229.7\\
EG-t      &\cred{2136.4}    &\cred{2167.8}  &2121.8         &\cred{2187.1}  &\cred{2239.3}  &2161.3         &2095.3         &2285.8         &2230.8\\
GJR-t     &2099.8           &2140.9         &2120.7         &2156.5         &2239.0         &\cred{2175.9}  &2077.6         &2287.6         &2230.3\\
Gt-HS     &2109.8           &2148.2         &2128.7         &2139.2         &2219.5         &2123.8         &2075.1         &2257.4         &2228.9\\
CARE      &2116.0           &2182.5         &2117.8         &2156.7         &2202.5         &2137.8         &\cred{2136.7}  &2232.6         &\bf{2321.1}\\
RG-RV-GG  &2093.5           &2146.0         &\cred{2217.4}  &2134.8         &2214.2         &2138.9         &2067.2         &\cred{2319.6}  &2202.9\\
RG-RV-tG  &2070.7           &2128.8         &2146.8         &2116.8         &2185.9         &2107.7         &2051.7         &2230.7         &2204.2\\
RC-Ra     &2107.7           &2131.2         &2118.8         &2141.4         &2193.2         &2129.0         &2093.0         &2230.0         &2222.4\\
RC-RaO    &2124.1           &2146.5         &2142.9         &2142.0         &2195.7         &2134.4         &2092.3         &2231.4         &2259.4\\
RC-RV     &2069.3           &2140.4         &2149.3         &2121.8         &2190.7         &\cb{2097.7}    &\fbox{2052.4}  &2229.3         &2214.5\\
RC-RR     &2060.2           &\fbox{2123.7}  &2128.1         &2120.1         &2177.4         &2099.0         &2068.5         &2226.1         &\fbox{2191.7}\\
RC-ScRV   &2078.0           &2142.2         &2124.0         &2123.8         &2196.8         &2101.1         &2056.1         &2231.9         &2215.4\\
RC-ScRR   &2066.8           &2133.7         &2111.7         &2118.7         &2182.8         &2095.0         &2073.2         &\fbox{2223.9}  &2208.1\\
RC-SubRV  &\fbox{2055.7}    &2130.7         &\cb{2110.2}    &2121.6         &\cb{2180.2}    &2095.8         &2058.4         &2226.9         &2201.0\\
RC-SubRR  &\cb{2057.0}      &2128.3         &2124.5         &2125.2         &\fbox{2180.1}  &2101.7         &2059.2         &\cb{2225.1}    &\cb{2194.2}\\
FC-Mean   &2071.6           &\cb{2126.0}    &\cb{2110.2}    &\fbox{2114.3}  &2182.6         &2098.7         &2053.6         &2233.1         &2212.7\\
FC-Med    &2069.3           &2130.5         &\fbox{2106.2}  &\cb{2116.2}    &2181.3         &\fbox{2096.7}  &\cb{2053.2}    &2230.2         &2208.6\\
FC-Min    &2123.2           &2144.6         &2139.5         &2146.0         &2201.2         &2132.6         &2101.6         &2237.7         &2250.9\\
FC-Max    &\bf{2174.9}      &\bf{2234.1}    &\bf{2253.2}    &\bf{2258.0}    &\bf{2303.8}    &\bf{2253.2}    &\bf{2194.2}    &\bf{2338.3}    &\cred{2293.7}\\
\hline
\end{tabular}
\end{center}
\emph{Note}:\small  Box indicates the favoured model, blue shading indicates the 2nd ranked model, bold indicates the least favoured model,
red shading indicates the 2nd lowest ranked model, in each column.
\end{table}

\begin{table}[!ht]
\begin{center}
\caption{\label{veloss)summary} \small VaR and ES joint loss function values summary; $\alpha=0.01$.}\tabcolsep=10pt
\begin{tabular}{lccccccccccccc} \hline
Model      & Mean loss          & Mean rank  \\ \hline
G-t        &2170.3          &14.67\\
EG-t       &\cred{2180.6}   &\cred{15.89}\\
GJR-t      &2169.8          &13.56\\
Gt-HS      &2158.9          &12.56\\
CARE       &2178.2          &14.33\\
RG-RV-GG   &2170.5          &12.22\\
RG-RV-tG   &2138.1          &6.67\\
RC-Ra      &2151.9          &9.89\\
RC-RaO     &2163.2          &13.56\\
RC-RV      &2140.6          &7.44\\
RC-RR      &2132.8          &4.44\\
RC-ScRV    &2141.1          &8.89\\
RC-ScRR    &2134.9          &4.89\\
RC-SubRV   &\fbox{2131.2}   &\fbox{3.67}\\
RC-SubRR   &2132.8          &5.00\\
FC-Mean    &2133.7          &5.33\\
FC-Med     &\cb{2132.5}     &\cb{4.11}\\
FC-Min     &2164.1          &14.00\\
FC-Max     &\bf{2255.9}     &\bf{18.89}\\
\hline
\end{tabular}
\end{center}
\emph{Note}:\small  Boxes indicate the favoured model, blue shading indicates the 2nd ranked model, bold indicates the least favoured model,
red shading indicates the 2nd lowest ranked model, in each column. "Mean rank" is the average rank across the 7 markets and 2 assets for the
loss function, over the 19 models: lower is better.
\end{table}

\subsubsection{\normalsize Model Confidence Set}

The model confidence set (MCS) was introduced by Hansen, Lunde and Nason (2011), as a method to statistically compare a
 group of forecast models via a loss function. We apply MCS to further compare among the 19 (VaR, ES) forecasting models.
A MCS is a set of models that is constructed such that it will contain the
best model with a given level of confidence, which was selected as 90\% in our paper. The Matlab code for MCS testing was
downloaded from "www.kevinsheppard.com/MFE\_Toolbox". We adapted code to incorporate the VaR and ES joint loss function values
(equation, \ref{eqveloss}) as the loss function during the MCS calculation. Each of two methods (R and SQ) to calculate the test statistics
are employed to the MCS selection process.

Table \ref{mcs_r} and \ref{mcs_sq} present the 90\% MCS using the R and SQ methods, respectively. Column "Total" counts the
total number of times that a model is included in the 90\% MCS across the 9 return series. Based on this column, boxes indicate the
favoured model, and blue shading indicates the 2nd ranked model for each market. Bold indicates the least favoured and red shading
indicates the 2nd lowest ranked model for each market.

Via the R method, Re-CARE-SubRV has the best performance and was included in the MCS for all markets and assets, followed by RG-RV-tG,
Re-CARE-SubRR and two forecasting combinations "FC-Mean" and "FC-Med", all included 8 times in the MCS in 9 series.
"FC-Max" is only included in the 90\% MCS twice, and Gt and Gt-HS are included only 3 times. Via the SQ method, the proposed
Re-CARE models are still favoured: the 90\% MCS includes Re-CARE-RR, Re-CARE-SubRV, Re-CARE-SubRR and RG-RV-tG in all 9 series,
followed by Re-CARE-ScRR, "FC-Mean" and "FC-Med", which are included 8 times.

\begin{table}[!ht]
\begin{center}
\caption{\label{mcs_r} \small 90\% model confidence set with R method across the markets and assets.}\tabcolsep=10pt
\tiny
\begin{tabular}{lccccccccccccc} \hline
Model    &S\&P 500  &NASDAQ &HK  &FTSE  &DAX  &SMI   &ASX200 &IBM &GE  &Total  \\ \hline
G-t      &0&1&0&0&0&1&0&1&0&\cred{3}\\
EG-t     &1&1&1&0&0&1&0&1&0&5\\
GJR-t    &1&1&1&1&0&1&1&0&0&6\\
Gt-HS    &0&1&0&1&0&1&0&0&0&\cred{3}\\
CARE     &0&1&1&0&1&1&0&1&0&5\\
RG-RV-GG &1&1&0&1&0&1&1&1&1&7\\
RG-RV-tG &1&1&0&1&1&1&1&1&1&\cb{8}\\
RC-Ra    &0&1&1&0&1&1&0&1&0&5\\
RC-RaO   &0&1&0&0&1&1&0&1&0&4\\
RC-RV    &1&1&0&1&1&1&1&1&0&7\\
RC-RR    &1&1&0&1&1&1&0&1&1&7\\
RC-ScRV  &1&1&0&1&0&1&1&1&0&6\\
RC-ScRR  &1&1&1&1&1&1&0&1&0&7\\
RC-SubRV &1&1&1&1&1&1&1&1&1&\fbox{9}\\
RC-SubRR &1&1&0&1&1&1&1&1&1&\cb{8}\\
FC-Mean  &1&1&1&1&1&1&1&1&0&\cb{8}\\
FC-Med   &1&1&1&1&1&1&1&1&0&\cb{8}\\
FC-Min   &0&1&0&0&1&1&0&1&0&4\\
FC-Max  &0&1&0&0&0&1&0&0&0&\bf{2}\\
\hline
\end{tabular}
\end{center}
\emph{Note}:\small  Boxes indicate the favoured model, blue shading indicates the 2nd ranked model, bold indicates the least favoured model,
red shading indicates the 2nd lowest ranked model, based on total number of included in the MCS across the 7 markets and 2 assets, higher is better.
\end{table}

\begin{table}[!ht]
\begin{center}
\caption{\label{mcs_sq} \small 90\% model confidence set with SQ method across the markets and assets.}\tabcolsep=10pt
\tiny
\begin{tabular}{lccccccccccccc} \hline
Model    &S\&P 500  &NASDAQ &HK  &FTSE  &DAX  &SMI   &ASX200 &IBM &GE  &Total  \\ \hline
G-t      &0&1&1&0&0&0&1&1&0&\cred{4}\\
EG-t     &0&1&1&0&0&1&1&1&0&5\\
GJR-t    &1&1&1&1&0&0&1&0&0&5\\
Gt-HS    &0&1&1&1&0&1&1&0&0&5\\
CARE     &0&1&1&0&1&0&0&1&0&\cred{4}\\
RG-RV-GG &0&1&0&1&0&1&1&1&1&6\\
RG-RV-tG &1&1&1&1&1&1&1&1&1&\fbox{9}\\
RC-Ra    &0&1&1&1&1&1&0&1&0&6\\
RC-RaO   &0&1&0&0&1&1&1&1&0&5\\
RC-RV    &1&1&0&1&1&1&1&1&0&7\\
RC-RR    &1&1&1&1&1&1&1&1&1&\fbox{9}\\
RC-ScRV  &0&1&1&1&0&1&1&1&0&6\\
RC-ScRR  &1&1&1&1&1&1&1&1&0&\cb{8}\\
RC-SubRV &1&1&1&1&1&1&1&1&1&\fbox{9}\\
RC-SubRR &1&1&1&1&1&1&1&1&1&\fbox{9}\\
FC-Mean  &1&1&1&1&1&1&1&1&0&\cb{8}\\
FC-Med   &1&1&1&1&1&1&1&1&0&\cb{8}\\
FC-Min   &0&1&0&0&1&1&0&1&0&\cred{4}\\
FC-Max   &0&0&0&0&0&0&0&0&0&\bf{0}\\
\hline
\end{tabular}
\end{center}
\emph{Note}:\small Boxes indicate the favoured model, blue shading indicates the 2nd ranked model, bold indicates the least favoured model,
red shading indicates the 2nd lowest ranked model, based on total number of included in the MCS across the 7 markets and 2 assets, higher is better.
\end{table}

Overall, across several measures and test for forecasts accuracy and model comparison, when forecasting 1\% VaR and ES in 9 financial
return series, the Re-CARE models employing RV and RR, and scaled and sub-sampled versions of those, generally performed in a highly
favourable manner when compared to a range of competing models. When considering VRates and rejections by standard tests, the Re-CARE-SubRR
model was the most favourable for VaR forecasting overall. When considering the joint loss function and bootstrap test, the
Re-CARE-SubRV model was the most favourable for ES forecasting overall. In each case, the best performing Re-CARE model also marginally
out-performed the forecast combination methods.

{\centering
\section{\normalsize CONCLUSION}\label{conclusion_section}
\par
}
\noindent
In this paper, the Realized-CARE, a new model framework to estimate and forecast financial tail risk, is proposed. Through incorporating
intra-day and high frequency volatility measures, improvements in the in-sample expectile estimation accuracy (compared to CARE model)
and out-of-sample forecasting of tail risk measures are observed, compared to Re-GARCH models employing realized volatility, and
traditional GARCH and CARE models, as well as forecast combinations of these models. Specifically, Re-CARE models with RaO, RR and
sub-sampled RR generate the most accurate VaR forecasts, while Re-CARE models employing RR, ScRV, SubRV are the most accurate
for ES forecasting in the empirical study of nine financial return series. Forecast combinations methods employing the mean and median of the
forecasts also produce competitive tail risk forecasting results. Regarding back testing of VaR forecasts, the Re-CARE type models are
also generally less likely to be rejected than their counterparts. Regarding the VaR and ES joint loss function values,
Re-CARE model's VaR and ES forecasts consistently had lower loss than all other models considered, especially the Re-CARE-RR and
Re-CARE-SubRV models. The combined series "FC-Mean" and "FC-Med" are also highly competitive regarding this loss function.
Further, the model confidence set results also favour the proposed Re-CARE framework, especially Re-CARE-RR, Re-CARE-SubRV and Re-CARE-SubRR.
In addition to being more accurate, the Re-CARE models generated less extreme tail risk forecasts, regularly allowing smaller amounts of
capital allocation without being anti-conservative or significantly inaccurate. The Re-CARE type
models with RR, sub-sampled RV, sub-sampled RR should be considered for financial applications when
forecasting tail risk, and should allow financial institutions to more accurately allocate capital under the Basel Capital Accord,
to protect their investments from extreme market movements. This work could be extended by developing different Re-CARE specifications,
perhaps considering alternative distributions for the measurement equation, by using alternative frequencies of
observation for the realized measures and extending the model to allow multiple realized measures and measurement equations, as per Hansen and Huang (2016).

\clearpage
\section*{References}
\addcontentsline{toc}{section}{References}
\begin{description}

\item Aigner, D.J. ,Amemiya, T., and Poirier, D. J. (1976). On the Estimation of Production
Frontiers: Maximum Likelihood Estimation of the Parameters of a Discontinuous Density
Function. \emph{International Economic Review}, 17, 377-396.

\item Andersen, T. G. and Bollerslev, T. (1998). Answering the skeptics: Yes, standard volatility models do provide accurate
forecasts. \emph{International economic review}, 885-905.

\item Andersen, T. G., Bollerslev, T., Diebold, F. X. and Labys, P. (2003). Modeling and forecasting realized volatility.
    \emph{Econometrica}, 71(2), 579-625.

\item Artzner, P., Delbaen, F., Eber, J.M., and Heath, D. (1997). Thinking coherently.  \emph{Risk}, 10, 68-71.

\item Artzener, P., Delbaen, F., Eber, J.M., and Heath, D. (1999). Coherent measures of risk.  \emph{Mathematical Finance}, 9, 203-228.

\item Bollerslev, T. (1986). Generalized Autoregressive Conditional Heteroskedasticity. \emph{Journal of Econometrics}, 31, 307-327.

\item Chang, C. L., Jim{\'e}nez-Mart{\'i}n, J. {\'A}., McAleer, M., and P{\'e}rez-Amaral, T. (2011). Risk management of risk under the
    Basel Accord: Forecasting value-at-risk of VIX futures.  \emph{Managerial Finance}, 37, 1088-1106.

\item Chen, Q, Gerlach, R. and Lu, Z. (2012). Bayesian Value-at-Risk and expected shortfall
forecasting via the asymmetric Laplace distribution. \emph{Computational Statistics and Data
Analysis}, 56, 3498-3516.

\item Christensen, K. and Podolskij, M. (2007). Realized range-based estimation of integrated variance. \emph{Journal of Econometrics},
141(2), 323-349.

\item Christoffersen, P. (1998). Evaluating interval forecasts. \emph{International Economic Review}, 39, 841-862.

\item Contino, C. and Gerlach, R. (2014). Bayesian Tail Risk Forecasting using Realised GARCH. The University of Sydney Business School
    working paper BAWP-2014-05 http://ses.library.usyd.edu.au/handle/2123/12060

\item Engle, R. F. (1982), Autoregressive Conditional Heteroskedasticity with Estimates of the Variance of United Kingdom
Inflations. \emph{Econometrica}, 50, 987-1007.

\item Engle, R. F. and Manganelli, S. (2004). CAViaR: Conditional Autoregressive Value at Risk
by Regression Quantiles. \emph{Journal of Business and Economic Statistics}, 22, 367-381.

\item Feller, W. (1951). The Asymptotic Distribution of the Range of Sums of Random Variables. \emph{Annals of Mathematical Statistics},
    22, 427-32.

\item Fissler, T. and Ziegel, J. F. (2016). Higher order elicibility and Osband's principle. \emph{Annals of Statistics}, in press.

\item Gaglianone, W. P., Lima, L. R., Linton, O. and Smith, D. R. (2011). Evaluating Value-
at-Risk models via quantile regression. \emph{Journal of Business and Economic Statistics},
29, 150-160.

\item Garman, M. B. and Klass, M. J. (1980). On the Estimation of Security Price Volatilities from historical data.
\emph{The Journal of Business}, 67-78.

\item Gerlach, R. and Chen, C.W.S. (2016). Bayesian Expected Shortfall Forecasting Incorporating the Intraday Range,
\emph{Journal of Financial Econometrics}, 14(1), 128–158.

\item Gerlach, R., Chen, C. W. S. and Lin, L. (2012). Bayesian Semi-parametric Expected Shortfall Forecasting in Financial Markets,
The University of Sydney working paper, available at http://ses.library.usyd.edu.au/handle/2123/8169.

\item Gerlach, R., Walpole, D. and Wang, C. (2016). Semi-parametric Bayesian Tail Risk Forecasting Incorporating Realized Measures of Volatility,
\emph{Quantitative Finance}, in press.

\item Gerlach, R. and Wang, C. (2016).  Forecasting risk via realized GARCH, incorporating the realized range.
\emph{Quantitative Finance}, 16:4, 501-511.

\item Gneiting, T (2011). Making and evaluating point forecasts. \emph{Journal of the American Statistical Association}, 106,
494, 746-762.

\item Hansen, B. E. (1994). Autoregressive conditional density estimation. \emph{International Economic Review}, 35, 705-730.

\item Hansen, P. R. and Huang, Z. (2016). Exponential GARCH Modeling
With Realized Measures of Volatility, \emph{Journal of Business \& Economic Statistics}, 34(2), 269-287,
DOI: 10.1080/07350015.2015.1038543.

\item Hansen, P. R., Huang, Z. and Shek, H. H. (2011). Realized GARCH: a joint model for returns and realized measures of volatility.
\emph{Journal of Applied Econometrics}, 27(6), 877-906.

\item Hansen, P.R., Lunde, A. and Nason, J.M. 2011. The model confidence set. Econometrica, 79(2), 453-497.

\item Hastings, W. K. (1970). Monte-Carlo Sampling Methods Using Markov Chains And Their Applications. \emph{Biometrika}, 57, 97-109.

\item Kupiec, P. H. (1995). Techniques for Verifying the Accuracy of Risk Measurement Models. \emph{The Journal of Derivatives}, 3, 73-84.

\item Martens, M. and van Dijk, D. (2007). Measuring volatility with the realized range. \emph{Journal of Econometrics}, 138(1), 181-207.

\item McAleer, M., Jim{\'e}nez-Mart{\'i}n, J. {\'A}. and P{\'e}rez-Amaral, T. (2013), GFC-robust risk management strategies under the
    Basel Accord,  \emph{International Review of Economics and Finance}, 27 , pp. 97-111.

\item Moln{\'a}r, P. (2012). Properties of range-based volatility estimators. \emph{International Review of Financial Analysis}, 23,
    20-29.

\item Metropolis, N., Rosenbluth, A. W., Rosenbluth, M. N., Teller, A. H., and Teller, E. (1953). Equation of State Calculations by Fast
Computing Machines. \emph{J. Chem. Phys}, 21, 1087-1092.

\item  Newey, W. K., and Powell, J.L. (1987). Asymmetric Least Squares Estimation and Testing. \emph{Econometrica}, 55, 819-847.

\item Parkinson, M. (1980). The extreme value method for estimating the variance of the rate of return. \emph{Journal of Business},
53(1), 61.

\item Rogers, L. C. G. and Satchell, S. E. (1991). Estimating variance from high, low and closing prices. \emph{Annals of Applied
    Probability}, 1, 504-512.

\item Roberts, G. O., Gelman, A. and Gilks, W. R. (1997). Weak convergence and optimal scaling of random walk Metropolis algorithms.
    \emph{The annals of applied probability}, 7(1), 110-120.

\item Schwert, G.W. (1989). Why Does Stock Market Volatility Change Over Time? \emph{Journal of Finance}, 44, 1115-1153.

\item So, M.K.P and Wong, C.M. (2011). Estimation of multiple period expected shortfall and median shortfall for risk management.
 \emph{Quantitative Finance}, 1-16.

\item Taylor, J. (2008). Estimating Value at Risk and Expected Shortfall Using Expectiles. \emph{Journal of Financial Econometrics}, 6,
    231-252.

\item Taylor, S.J. (1986). Modeling Financial Time Series. Chichester, UK: John Wiley and Sons.

\item Watanabe, T. (2012). Quantile Forecasts of Financial Returns Using Realized GARCH Models. \emph{Japanese Economic Review}, 63(1),
    68-80.

\item Yang, D. and Zhang, Q. (2000). Drift-independent volatility estimation based on high, low, open, and close prices. \emph{Journal of
    Business}, 73, 477-491.

\item Zhang, L., Mykland, P. A., and A\"{i}t-Sahalia, Y. (2005). A tale of two time scales.  \emph{Journal of the American Statistical
    Association}, 100(472).

\end{description}

\end{document}